\documentclass[lineno]{jfm}
\usepackage{graphicx, float}
\usepackage{amsmath}
\usepackage{mathrsfs}
\usepackage[utf8]{inputenc}
\usepackage{natbib}
\usepackage{caption}
\usepackage[position=top]{subcaption}
\usepackage{comment}
\usepackage{bbding}
\usepackage[export]{adjustbox}
\usepackage{standalone}
\usepackage{rotating}
\usepackage[ruled,vlined]{algorithm2e}
\usepackage[T1]{fontenc}

\usepackage[per-mode=symbol]{siunitx}
\usepackage[normalem]{ulem}
\usepackage{cancel}
\usepackage{booktabs}
\usepackage{threeparttable}
\usepackage{booktabs}
\usepackage{array}
\usepackage{makecell}

\newcolumntype{L}[1]{>{\raggedright\arraybackslash}m{#1}}
\newcolumntype{C}[1]{>{\centering\arraybackslash}m{#1}}
\usepackage{booktabs}
\usepackage{array}
\usepackage{adjustbox}
\usepackage{booktabs}
\usepackage{array}
\usepackage{adjustbox}

\newcommand{\blcorr}{\mathcal{B}_n}
\newcommand{\etaIrrot}{\frac{2(n+1)(n+2)\mu}{\rho R_o^2}}
\newcommand{\etaApprox}{\frac{2(2n+1)(n+2)\mu}{\rho R_o^2}}
\newcommand{\etaBL}{\frac{2(n+2)\mu}{\rho R_o^2}\blcorr}
\newcommand{\xiCurrentBL}{\frac{2(n+2)}{\rho R_o^2}\left[\blcorr G+\frac{(n+1)(n-1)\gamma}{2R_o}\right]}
\newcommand{\xiPlesset}{\frac{(n+1)(n+2)(n-1)\gamma}{\rho R_o^2}}

\newcommand{\xiRemillard}{\frac{2(n+2)(n+1)}{\rho R_o^2}\left[G+\frac{(n-1)\gamma}{2R_o}\right]}
\newcommand{\xiCurrentApprox}{\frac{2(n+2)}{\rho R_o^2}\left[(2n+1)G+\frac{(n+1)(n-1)\gamma}{2R_o}\right]}

\usepackage[dvipsnames]{xcolor}
\usepackage{pdflscape}
\usepackage[hidelinks]{hyperref}
\usepackage[nameinlink,noabbrev]{cleveref}
\crefname{equation}{equation}{equations}
\usepackage{upgreek}
\usepackage{bm}
\usepackage{amsmath,amssymb,mathtools,bm}
\usepackage{physics}
\usepackage{microtype}
\usepackage{tikz}
\usetikzlibrary{decorations.pathreplacing}
\usepackage{breakcites}
\usetikzlibrary{shapes}
\usepackage{mathtools}
\usepackage{lipsum}
\definecolor{red2}{RGB}{192, 16, 16}
\definecolor{green2}{RGB}{48, 128, 48}
\definecolor{green3}{RGB}{0, 200, 0}
\definecolor{blue2}{RGB}{16, 80, 160}
\definecolor{blue3}{RGB}{0, 0, 255}
\definecolor{orange2}{RGB}{192, 128, 64}
\definecolor{yellow2}{RGB}{224, 192, 32}
\definecolor{purple2}{RGB}{96, 64, 160}
\definecolor{teal2}{RGB}{32, 112, 144}

\newcommand{\T}{\mathsf{T}}
\usepackage{xcolor}
\usepackage{multirow}
\definecolor{drkg}{RGB}{105,105,105}
\definecolor{midg}{RGB}{169,169,169}
\definecolor{lgtg}{RGB}{220,220,220}
\definecolor{drkb}{RGB}{0,25,51}
\definecolor{lgtb}{RGB}{0,102,204}
\definecolor{lightg}{RGB}{160,160,160}
\definecolor{lightgr}{RGB}{30,120,30}
\newcommand\Ca{\mbox{\textit{Ca}}}
\newcommand\We{\mbox{\textit{We}}}
\newcommand\Oh{\mbox{\textit{Oh}}}
\newcommand\Ec{\mbox{\textit{Ec}}}

\usepackage{color}

\newcommand{\YY}{Y_n^m}

\title{Nonspherical gas bubble dynamics in viscoelastic soft materials}

\shorttitle{Nonspherical gas bubble dynamics in viscoelastic soft materials}

\author{Sawyer Remillard \and
  Mauro Rodriguez Jr.\corresp{\email{mauro\_rodriguez@brown.edu}} }

\shortauthor{S. Remillard and M. Rodriguez Jr.}

\affiliation{
School of Engineering, Brown University, 
Providence, RI 02912, USA
}

\begin{document}

\maketitle
\nolinenumbers
\begin{abstract}

Nonspherical gas bubble dynamics in viscoelastic materials influence the stress transmission and energy dissipation of their surroundings and are difficult to predict.
Their accurate prediction is essential in applications ranging from biomedical procedures to high-strain-rate rheological measurements.
However, existing models do not sufficiently capture the nonspherical rotational dynamics.
We formulate and superpose a rotational contribution to the perturbed deformation with a potential contribution.
Linearised forward and inverse coordinate maps are formulated based on the deformation field which are used to compute velocities, accelerations, and stresses.
The addition of the rotational degree of freedom satisfies the momentum balance equations and stress continuity at the bubble surface.
The material surrounding the bubble is modelled with a Kelvin-Voigt constitutive model with Newtonian viscosity and quadratic strain-stiffening neo-Hookean elasticity. 
The model agrees with previous viscous fluids models when elastic effects are neglected and radial oscillations are small.
When viscous effects are small relative to elastic, shear waves radiate from the bubble surface into the material.
The resulting strain energy is delocalised and increases damping of the perturbation amplitude in time relative to potential-based models.
We show agreement between the stability of the shape modes with previous ultrasound forced experiments and temporal evolution of different shape modes with previous laser-induced cavitation experimental data.

\end{abstract}


\section{Introduction}

Instabilities in spherical physical systems involving viscous, elastic, and viscoelastic material surroundings are difficult to predict and control. 
In self-gravitating media, pressure or elastic-like restoring stresses oppose gravitational collapse in Jeans-type instabilities \citep{Jeans1902,Bonnar1956,Janaki2011}, thereby setting critical length and mass scales for gravitational instability and cloud fragmentation \citep{Hoyle1953,McKee2007}. 
Similarly, Bell--Plesset effects amplify the evolution of the Rayleigh-Taylor Instability (RTI) in spherical geometries \citep{Li2026}, which may explain mixing in core-collapse supernovae \citep{Burrows2000,Gamezo2003,Joggerst2010,Yadav2020}.
Similarly, these instabilities in convergent geometries limit energy concentration in inertial confinement fusion (ICF) experiments \citep{MacPhee2017,Ali2018} by enhancing interface mixing \citep{Ramshaw1998,Ramshaw1999,Ramshaw2003}.
The key system of interest is nonspherical cavitation-bubble collapse which can subsequently damage nearby materials and surfaces.

Cavitation bubbles become nonspherical due to their surroundings.
These nonspherical surface perturbations are generated due to, for example, non-uniform pressure fields (e.g., ultrasound forcing \citep{Dollet2008,Versluis2010,Lajoinie2018,ALPACASim} or shock-driven \citep{KLASEBOER2007,Johnsen2008}), and/or influence from other nearby bubbles or boundaries \citep{Blake1993,Bremond2006,Plesset1971,BRUJAN2001,Lauer2012}. 
Near rigid flat walls, nonspherical bubbles can generate a liquid re-entrant jet that impinge on nearby material surfaces and generate localised damage \citep{Blake1987,PHILIPP1998,Supponen2016,JOHNSEN_COLONIUS_2009,Supponen2017,Rodriguez2022}.
These cavitation events limit propeller and turbomachinery life cycles \citep{Melissaris2023,Usta2023} and the efficacy of ultrasound contrast agents \citep{Dollet2008,Vos2011} and focused-ultrasound therapy tools, e.g., histotripsy \citep{Mancia2020,Bader2019,Parsons2006,Vlaisavljevich2015} and lithotripsy \citep{Bailey2005,Zeman1990}. 
Moreover, controlling these perturbations is necessary to drive payload release from targeted microbubbles \citep{Lajoinie2018} and in sonoporation \citep{Ohl2006,Wang2018,Rich2021}.
These perturbations enhance ultrasound-forcing and Laser-Induced Cavitation (LIC) rheology of soft viscoelastic materials \citep{Remillard2026}; and may constrain needle-induced cavitation rheology \citep{Barney2020,Zimberlin2007,RaayaiArdakani2019}.

Bubble interface perturbations have been investigated using primarily theoretical approaches.
The first studies primarily investigated the stability of interface perturbations when subjected to bubble growth or collapse \citep{Binnie1953}, and the nonspherical oscillation frequency \citep{Lamb,Strasberg1953}.
Linearised models with respect to the nonspherical motion have been proposed to predict the evolution of the surface instabilities.
For example, in inviscid materials \citep{Plesset1954}, viscous fluids \citep{Prosperetti1977}, encapsulated within a thin elastic layer \citep{Matsumoto_2012}, Maxwell-type viscoelastic materials \citep{Hara1984}, and Kelvin-Voigt-type materials \citep{Gaudron2020,Murakami2020,Yang2021,Remillard2026}.
Models which retain higher orders of the nonspherical motion account for the energy transfer between the shape and volume modes which introduces a two-way coupling between the radial and nonspherical oscillations in fluids \citep{Shaw2006,Shaw2025Pof,Shaw2025JMF,Claude2017,Harkin2013}.
Despite assuming irrotational flow, these models recover the damping coefficient predicted by \citet{Lamb} when $\dot{R}=0$, unlike linearised potential-based models.

Theoretical nonspherical bubble models should satisfy mass conservation, momentum balance, and stress continuity at the bubble surface.
If the bubble is in a free field, the stress and kinematic equilibrium should also be satisfied in the far field.
For a bubble surrounded by an inviscid fluid, a purely potential description of the velocity (potential flow) is sufficient because the surrounding material cannot sustain shear stress \citep{Plesset1954}, thus tangential stress continuity is trivially satisfied.
Since the flow is irrotational, the vorticity equation is also trivially satisfied. 
However, viscous, elastic, or viscoelastic materials, which experience shear stresses, generate non-trivial conditions for tangential stress continuity at the bubble interface and vorticity.
In viscous liquids, \citet{Prosperetti1977} solved these conditions by adding a rotational degree of freedom in the kinematics to the \citet{Plesset1954} model.
To compute finite elastic stresses, it is necessary to define the kinematics in terms of deformation.
However, previous potential \citep{Remillard2026} or purely radial \citep{Gaudron2020,Murakami2020,Yang2021} deformation models do not satisfy all components of the momentum balance.
Although the model from \cite{Hara1984} includes the necessary rotations, it is limited to small radial deformations.
Additionally, Maxwell-type stress relaxation effects are typically on much longer bubble oscillation length and time scales.
Bubbles within soft viscoelastic materials are better described by a Kelvin-Voigt-type model \citep{Gaudron2015,Hamaguchi2015,Mohaved2016,Chu2025,Sanchez2026}.

We formulate a theoretical model for nonspherical bubble dynamics in a Kelvin-Voigt-type material with hyperelasticity and validate it with experimental data. 
The model satisfies the momentum balance equations and tangential projection of stress continuity across the bubble surface by including the kinematic rotational degrees of freedom.
In the context of ultrasound forced and laser-induced nonspherical oscillations, we investigate two types of instabilities predicted by the model: parametric and the classic Rayleigh-Taylor instability.
The paper is organised as follows.
The problem is formulated in \cref{sec:problem_formulation}.
The solution is a sum of spherical and nonspherical potential and rotational contributions in~\cref{sec:solution_strategy}.
The equations governing the surface perturbations and spatial and temporal evolution of the rotational components of the deformation field are formulated in~\cref{sec:governing_equations}.
The problem setup and numerical methods to solve the governing equations are detailed in~\cref{sec:problem_setup}.
In~\cref{sec:results}, we compare and discuss previous models and experimental data with the present full model results for small and finite radial oscillations.
For small oscillations, we quantify the rotation-based elastic dissipation due to radiative shear wave emission.
We then present the stability of the finite radial oscillations under ultrasonic forcing and nonspherical mode and stress evolution under LIC loading.
We summarise our contributions in \cref{sec:conclusions}.

\section{Problem formulation \label{sec:problem_formulation}}
We consider a nonspherical gas bubble in sea of an incompressible viscoelastic material.
The bulk radial motion can be finite, and the interface perturbations are small.
The problem is formulated in the respective spherical referential and current coordinates and position vectors: 
\begin{subequations}
    \begin{equation}
    \psi_o = \left(r_o, \theta_o, \phi_o\right), \quad \psi = \left(r, \theta, \phi \right),
\end{equation}
\begin{equation}
    \mathbf{x}_o = r_o \, \hat{\mathbf{e}}_r(\theta_o, \phi_o), \quad \mathbf{x} = r \, \hat{\mathbf{e}}_r(\theta, \phi),
    \label{eq:position}
\end{equation}
\end{subequations}
where $\mathbf{\hat{e}}_r$ is the radial direction unit vector. 
Quantities with subscript $o$ are in the reference coordinates.
The bubble surface is defined by the level set:
\begin{equation}
    \Sigma(t) : F(r, \theta, \phi,t) \equiv r-\mathscr{R}(t, \theta, \phi)=r-R(t)\left(1+\sum_{n,m}\epsilon_n(t) \YY(\theta, \phi)\right),
    \label{eq:bub_surf}
\end{equation} 
and is zero at the bubble surface.
$R$ is the mean bubble radius, $\epsilon_n$ the perturbation amplitude of a given mode $n$, and $Y_n^m$ the real spherical harmonics.
To first order in the perturbation amplitude, the equations for each mode are uncoupled from one another and independent of order $m$.
For brevity, the summation and time dependency notation of $R$ and $\epsilon_n$ are dropped.
The equations of motion for the surrounding material are
\begin{subequations}
    \begin{equation}
        \nabla_{\mathbf{x}}\cdot\mathbf{v} = 0,
    \end{equation}
    \begin{equation}
        \frac{\partial \mathbf{v}}{\partial t} + \left(\mathbf{v}\cdot\nabla_{\mathbf{x}} \right)\mathbf{v} = \frac{1}{\rho}\left(\nabla_{\mathbf{x}}\cdot\boldsymbol{\sigma} \right),
        \label{eq:mom_balance}
    \end{equation}
\end{subequations}
where $\nabla_{\mathbf{x}}$ is the current coordinate gradient operator, $\mathbf{v}$ the Eulerian velocity vector, $\rho$ the surrounding material density, and $\boldsymbol{\sigma}$ the Cauchy stress tensor.
The constitutive response of the material is modelled with the quadratic strain stiffening Kelvin-Voigt model with neo-Hookean elasticity and Newtonian viscosity with elastic strain energy density function, 
\begin{equation}
    \psi  = \frac{G}{2}\left[ \left(\mathrm{I}_{\mathbf{B}}-3 \right)+\frac{\alpha}{2}\left(\mathrm{I}_{\mathbf{B}}-3 \right)^2\right],
    \label{eq:free_energy_density}
\end{equation}
and Cauchy stress,
\begin{equation}
    \boldsymbol{\sigma} = 
    \begin{cases}\boldsymbol{\sigma}_\text{m} = G\left(1+\alpha(\mathrm{I}_{\mathbf{B}}-3) \right)\mathbf{B}+2\mu \mathbf{D}-\mathscr{P}\mathbf{I}, & \text { if } r>\mathscr{R}, \\ 
    \boldsymbol{\sigma}_\text{b} = -p_\text{b}\mathbf{I} & \text { otherwise }.
    \end{cases}
\end{equation}
The respective subscripts m and b denote the surrounding material and bubble, $G$ is the shear modulus, $\alpha$ the strain stiffening parameter, $\mathbf{B} = \mathbf{F F}^{\textrm{T}}$ the left Cauchy-Green deformation tensor, $\mathrm{I}_{\mathbf{B}}$ the first invariant of $\mathbf{B}$, $\mathbf{F}$ the deformation gradient tensor, $\mu$ the viscosity, $p_{\rm b}$ the bubble pressure, and $\mathbf{I}$ the identity tensor.
$\mathscr{P}$ is the Lagrange multiplier enforcing incompressibility and $\mathbf{D}$ the rate of strain tensor:
\begin{equation*}
    \mathbf{D} = \frac{1}{2}\left(\nabla_{\mathbf{x}}\mathbf{v}+(\nabla_{\mathbf{x}}\mathbf{v})^{\text{T}} \right).
\end{equation*}
Continuity of tangential deformation across the bubble surface is not considered since the gas inside the bubble is approximately inviscid \citep{Prosperetti1977,Murakami2020thesis}. 

To satisfy the tangent and normal stress continuity at the bubble surface:
\begin{subequations}
    \begin{equation}
        \mathbf{n}\times\left[\left(\boldsymbol{\sigma}_\text{m}-\boldsymbol{\sigma}_\text{b} \right)\mathbf{n} \right] = \mathbf{0}, 
        \label{eq:tang_stress_continuity_BC}
    \end{equation}
    \begin{equation}
        \mathbf{n}\cdot\left[\left(\boldsymbol{\sigma}_\text{m}-\boldsymbol{\sigma}_\text{b} \right)\mathbf{n} \right] = \gamma\nabla_{\mathbf{x}}\cdot\mathbf{n}, 
        \label{eq:norm_stress_continuity_BC}
    \end{equation}
\end{subequations}
respectively, where $\gamma$ is the surface tension between the gaseous bubble and surrounding material. 
$\mathbf{n}$ is the outward pointing normal on $\Sigma$:
\begin{equation}
    \mathbf{n} = \mathbf{e}_r-\epsilon_n\frac{R}{r}\frac{\partial}{\partial\theta}\YY\mathbf{e}_{\theta} - \epsilon_n\frac{R}{r \sin \theta}\frac{\partial}{\partial\phi}\YY \mathbf{e}_{\phi}.
\end{equation}
The material stress equilibrates in the far field:
\begin{equation}
    \boldsymbol{\sigma}(r\to \infty) = -p_{\infty}\mathbf{I},
    \label{eq:stress_equil_infty}
\end{equation}
where $p_{\infty}$ is the far-field pressure.

Following \citet{Prosperetti1977}, we seek deformation and pressure solutions: 
\begin{subequations}
    \begin{equation}
        \mathbf{u} = \mathbf{u}_s+\varepsilon\mathbf{u}_{\text{p}}+\varepsilon\mathbf{u}_{\text{rot}},
    \end{equation}
    \begin{equation}
        \mathscr{P} = \mathscr{P}_s+\varepsilon\mathscr{P}_{\text{p}}+\varepsilon\mathscr{P}_{\text{rot}}, 
    \end{equation}
\end{subequations}
where terms with subscripts $s$, p, and rot are the purely spherical, the potential, and rotational contributions, respectively, and $0<\varepsilon \ll 1$. 
Pressure $\mathscr{P}$ is used to satisfy \cref{eq:norm_stress_continuity_BC,eq:stress_equil_infty}.
The current position vector is decomposed into the base radial deformation with superposed first order small nonspherical deformation, 
\begin{equation}
    \mathbf{x} = r_s\hat{\mathbf{e}}_r(\theta_o,\phi_o)+\varepsilon\mathbf{u}_1 + \mathscr{O}(\varepsilon^2),
    \label{eq:small_on_large}
\end{equation}
where $r_s$ is the base spherical deformation and $\mathbf{u}_1$ the deformation field due to nonspherical surface perturbations.
Conventionally, $\mathbf{u}$ is the displacement vector; in the absence of rigid body motion it is the deformation field.
Since the nonspherical motion is small relative to the spherical motion, terms $\mathscr{O}(\varepsilon^2)$ and higher are truncated.
Then, the respective maps between the reference and current coordinates are
\begin{subequations}
    \begin{equation}
        \psi = \chi\left(\psi_o \right) \approx \chi_s\left( \psi_o\right) + \varepsilon\chi_1\left(\psi_o \right),
        \label{eq:map}
    \end{equation} 
    \begin{equation}
        \psi_o=\Upsilon(\psi) \approx \Upsilon_s(\psi)+\varepsilon\Upsilon_1(\psi),
    \label{eq:inv_map}
    \end{equation}
\end{subequations}
where $\chi$ and $\Upsilon$ are the forward and inverse coordinate maps.
The respective base state contributions to the total maps are 
\begin{equation*}
    \begin{aligned}
        \chi_s\left( \psi_o\right)& =  \left(r_s(r_o, t), \, \theta_o, \, \phi_o\right),\\
        \Upsilon_s\left( \psi\right)& =  \left(\varrho_s(r, t), \, \theta, \, \phi\right),
    \end{aligned}
\end{equation*}
where $\varrho$ is the radial component to the inverse map.
Similarly, the respective linear contributions are
\begin{equation*}
    \begin{aligned}
        \chi_1\left( \psi_o\right) &= \left(r_1(\psi_o, t), \,  \theta_1(\psi_o, t),\,  \phi_1(\psi_o, t)\right), \\
        \Upsilon_1\left( \psi\right) &= \left(\varrho_1(\psi, t), \,  \vartheta_1(\psi, t),\,  \varphi_1(\psi, t)\right).
    \end{aligned}
\end{equation*}
The linear contributions are decomposed into a potential solution with the rotational correction: $(\cdot)_1 = (\cdot)_{\textrm{p}} + (\cdot)_{\textrm{rot}}$.
Using the coordinate maps and \cref{eq:position,eq:small_on_large}, the linearised deformation field:
\begin{equation}
    \mathbf{u}_1 = r_1\hat{\mathbf{e}}_r + r_s\theta_1\hat{\mathbf{e}}_\theta + r_s\sin\theta_o \phi_1 \hat{\mathbf{e}}_\phi,
    \label{eq:pert_disp}
\end{equation}
where $\hat{\mathbf{e}}_\theta$ and $\hat{\mathbf{e}}_\phi$ are unit vectors in the polar and azimuthal direction, respectively.
The unit vectors in \cref{eq:pert_disp} are functions of $\theta_o$ and $\phi_o$.
Since $(\theta,\phi) = (\theta_o, \phi_o)+\mathscr{O}(\varepsilon)$, then the unit vectors of $\mathbf{u}_1$ can be written in either current or reference coordinate basis to first order accuracy in $\varepsilon$.

The deformation gradient can be decomposed as a sum of powers of $\varepsilon$,
\begin{equation}
    \mathbf{F} = \mathbf{F}_s+\varepsilon \mathbf{F}_1.
\end{equation} 
Truncating $\mathscr{O}(\varepsilon^2)$ and higher terms, the deformation gradient tensor is in \citet{Remillard2026}.
Material incompressibility is satisfied with $\det(\mathbf{F}) = 1$.
After linearising, the incompressibility condition becomes
\begin{equation*}
    \text{det}(\mathbf{F}) \approx \text{det}(\mathbf{F}_s)+\varepsilon \left[\frac{\partial}{\partial \varepsilon} \text{det}(\mathbf{F})\right]_{\varepsilon = 0} = 1.  
\end{equation*}
Thus, the incompressibility condition for the perturbed deformation field yields
\begin{equation}
    \left[\frac{\partial}{\partial \varepsilon} \text{det}(\mathbf{F})\right]_{\varepsilon = 0}  = \text{tr}\left(\mathbf{F}_s^{-1}\mathbf{F}_1 \right)=\nabla_{\mathbf{x}}\cdot (\mathbf{u}_{\text{p}}+\mathbf{u}_{\text{rot}})=0.
    \label{eq:lin_incomp}
\end{equation}
Since $\mathbf{u}_{\rm p}$ is divergence free~\citep{Remillard2026}, so is $\mathbf{u}_{\text{rot}}$.

The Eulerian-Almansi strain tensor is used to describe strains in the current configuration:
\begin{equation}
    \mathbf{e} = \frac{1}{2}\left(\mathbf{I}-\mathbf{B}^{-1} \right) \approx \mathbf{e}_s+\varepsilon\mathbf{e}_1,
\end{equation}
where
\begin{subequations}
    \begin{equation}
        \mathbf{e}_s = \frac{1}{2}(\mathbf{I}-\mathbf{B}_s^{-1}),
    \end{equation}
    \begin{equation}
        \mathbf{e}_1 = \frac{1}{2}\left( (\nabla_{\mathbf{x}}\mathbf{u}_1)^{\rm T}\mathbf{B}_s^{-1}+\mathbf{B}_s^{-1} \nabla_{\mathbf{x}}\mathbf{u}_1 \right),
    \label{eq:small_strain}
    \end{equation}
\end{subequations}
with $\mathbf{B}_s^{-1} = \mathbf{F}_s^{-\rm T}\mathbf{F}_s^{-1}$.
If the base spherical motion is static or linear, i.e. $\mathbf{B}_s \approx \mathbf{I}$,  \cref{eq:small_strain} simplifies to the small strain tensor.

\section{Solution of the viscoelastic continuum\label{sec:solution_strategy}}
\subsection{Spherical and potential deformation field solution}

The base state deformation gradient tensor for the purely spherical case is  \citep{Gaudron2015}
\begin{equation}
    \mathbf{F}_s=\left[\begin{array}{ccc}
\frac{\partial r_s}{\partial r_o} & 0 & 0\\
0 & \frac{r_s}{r_o} & 0 \\
0 & 0 & \frac{r_s }{r_o} 
\end{array}\right].
\end{equation}
Equating the determinant of $\mathbf{F}_s$ to unity, the incompressibility relationship for the spherical forward and inverse map is
\begin{subequations}
\begin{equation}
    r_s  = \left(r_{o}^3+ R^3 -R_o^3\right)^{1/3},
    \label{eq:sph_map}
\end{equation}
\begin{equation}
    \varrho_s=\left(r^3-R^3+R_o^3\right)^{1/3}.
    \label{eq:sph_inv_map}
\end{equation}
\end{subequations}
where $R_o$ is the mechanical equilibrium radius in the referential coordinate system.
The deformation field formulation follows that of \citet{Prosperetti1977} for the velocities.
Thus, we seek the general solution to satisfy the boundary conditions and equations of motion.
The potential deformation field solution~\citep{Remillard2026}:
\begin{equation}
    \mathbf{u}_{\rm p} = -\epsilon_n\frac{R^{n+3}}{(n+1)}\nabla_{\mathbf{x}}\left(\frac{\YY}{r^{n+1}}\right),
    \label{eq:potential_displacments}
\end{equation}
satisfies the radial deformation condition $\left.r_1\right|_{r_o = R_o} = \epsilon_nR\YY$ ensuring the level set (\cref{eq:bub_surf}) is zero at the bubble surface after applying the forward map.

\subsection{Rotational correction}
The rotation of the deformation field is defined as
\begin{equation}
    \boldsymbol{\omega} = \nabla_{\mathbf{x}}\times \mathbf{u} = \nabla_{\mathbf{x}}\times \mathbf{u}_{\text{rot}},
\end{equation}
where 
\begin{subequations}
    \begin{equation}
        \nabla_{\mathbf{x}}\cdot\mathbf{u}_{\text{rot}} = 0,
        \label{eq:rot_incomp}
    \end{equation}
    \begin{equation}
        \nabla_{\mathbf{x}}\cdot\boldsymbol{\omega} = 0.
    \end{equation}
\end{subequations}
The deformation field rotation is decomposed into a sum of poloidal and toroidal contributions:
\begin{equation}
    \boldsymbol{\omega} = \nabla_{\mathbf{x}}\times\nabla_{\mathbf{x}}\times\left[S_n(r,t)\YY \hat{\mathbf{e}}_r \right] + \nabla_{\mathbf{x}}\times\left[T_n(r,t)\YY \hat{\mathbf{e}}_r \right],
\end{equation}
with corresponding deformation field:
\begin{equation}
    \mathbf{u}_{\text{rot}} = T_n(r,t)\YY \hat{\mathbf{e}}_r +\nabla_{\mathbf{x}}\times\left[S_n(r,t)\YY \hat{\mathbf{e}}_r \right] -\nabla_{\mathbf{x}}[\Phi_n(r,t)\YY].
    \label{eq:rot_disp}
\end{equation}
The final term in \cref{eq:rot_disp} has been added to enforce incompressibility: 
\begin{equation}
    \nabla_{\mathbf{x}}^2(\Phi_n(r,t)\YY) = \nabla_{\mathbf{x}}\cdot\left(T_n(r,t)\YY \hat{\mathbf{e}}_r \right),
    \label{eq:rot_incomp2}
\end{equation}
which can be solved using Green's functions.
The potential solution satisfies the radial deformation boundary condition, thus the radial component of $\mathbf{u}_{\textrm{rot}}$ is zero at the bubble wall.
The radial component of the velocity boundary condition from \citet{Prosperetti1977} for the rotational field (i.e., zero at the bubble wall and in the far field) is analogous to the deformation boundary conditions.
Using their general solution for $\Phi_n$:
\begin{equation}
    \Phi_n(r,t) =  \frac{n+1}{2n+1}\mathcal{K}_n(r,t)r^{n}+\left(\frac{n}{n+1} R^{2n+1}\kappa_n(t)+ \frac{n}{2n+1}\mathcal{J}_n(r,t)\right) r^{-(n+1)},
\end{equation}
where 
\begin{subequations}
   \begin{equation}
       \mathcal{J}_n(r,t) = \int_R^r s^{n+1} T_n(s, t) d s,
   \end{equation} 
   \begin{equation}
       \mathcal{K}_n(r,t) = -  \int_r^\infty s^{-n} T_n(s, t) d s,
   \end{equation}
   \begin{equation}
       \kappa_n(t) = -\frac{n+1}{2n+1}\int_{R}^{\infty} s^{-n} T_n(s,t) ds.
   \end{equation}
\end{subequations}
The forward maps can be solved using \cref{eq:pert_disp,eq:potential_displacments,eq:rot_disp,eq:map}.
To obtain the inverse coordinate maps, we invert the forward map by orders of $\varepsilon$.
To determine $\Upsilon_1$, the forward map is written component wise as $\psi_i=\chi_i(\psi_o,t,\varepsilon)$, and the residuals are defined as
\begin{equation}
    \mathcal{F}_i(\psi_o,\psi,t,\varepsilon) \equiv \chi_i(\psi_o,t,\varepsilon)-\psi_i.
\end{equation}
The inverse map is obtained by setting these residuals to zero to first order in $\varepsilon$ after substituting $\psi_o=\Upsilon_s(r)+\varepsilon\Upsilon_1(\psi)$.
Expanding the residuals about $\varepsilon=0$ gives
\begin{equation}
    \left.\nabla_{\psi_o}\mathcal{F}\right|_{\varepsilon=0,\psi_o=\Upsilon_s}\Upsilon_1=-\left.\frac{\partial \mathcal{F}}{\partial \varepsilon}\right|_{\varepsilon=0,\psi_o=\Upsilon_s},
    \label{eq:inv_map_eq}
\end{equation}
which is solved to find the three linear contributions to the inverse map $(\varrho_1,\vartheta_1,\varphi_1)$.
The forward and inverse maps are summarised in \cref{sec:appendix_maps}, and velocities and accelerations, 
\begin{subequations}
    \begin{equation}
        \mathbf{v} = \dot{r} \, \hat{\mathbf{e}}_r + r\dot{\theta} \, \hat{\mathbf{e}}_{\theta}+ r\sin\theta \dot{\phi} \, \hat{\mathbf{e}}_{\phi},
        \label{eq:velocity}
    \end{equation}
    \begin{equation}
        \begin{aligned}
            \boldsymbol{a} &= \left(\ddot{r}-r\dot{\theta}^2-r\dot{\phi}^2  \sin^2\theta \, \right)\hat{\mathbf{e}}_r + \left(r\ddot{\theta}+ 2\dot{r}\dot{\theta}-r\dot{\phi}^2\sin\theta \cos\theta \right) \hat{\mathbf{e}}_{\theta} \\
            &+ \left(r\ddot{\phi}  \sin \theta +2\dot{r}\dot{\phi}\sin\theta+ 2 r \dot{\theta}\dot{\phi}\cos \theta\right) \hat{\mathbf{e}}_{\phi}, 
        \end{aligned}
    \end{equation}
\end{subequations}
where dot operators are material (substantial) time derivatives.
Although we use the same solution method as \cite{Prosperetti1977}, when $\dot{R}\neq 0$, the time derivative of the deformation field is not generally the same as the kinematic space used by \citet{Prosperetti1977} for Eulerian velocity fields.

\section{Evolution equations\label{sec:governing_equations}}
\subsection{Field equations and interface boundary conditions}
To close \cref{eq:mom_balance} through the scalar pressure field $\mathscr{P}$, the vorticity equation must be satisfied.
Taking the curl of \cref{eq:mom_balance} and applying spherical harmonic orthogonality, the respective governing equations for the toroidal and poloidal fields are:
\begin{subequations}
    \begin{equation}
    \begin{aligned}
        &\frac{\partial^2 T_n}{\partial t^2} + \mathscr{C}_{t1}\frac{\partial^3 T_n}{\partial r^3}+ \mathscr{C}_{t2}\frac{\partial^3 T_n}{\partial t\partial r^2}+\mathscr{C}_{t3}\frac{\partial^2 T_n}{\partial r^2}+\mathscr{C}_{t4}\frac{\partial^2 T_n}{\partial t\partial r} + \\& \mathscr{C}_{t5}\frac{\partial T_n}{\partial r} + \mathscr{C}_{t6}\frac{\partial T_n}{\partial t}  + \mathscr{C}_{t7}T_n+ \mathscr{G}_{\Phi_n}(T_n)+ \mathscr{G}_{\rm p}(\epsilon_n) = 0,
        \end{aligned}
        \label{eq:T_bulk}
    \end{equation}
    \begin{equation}
        \begin{aligned}
            &\frac{\partial^2 S_n}{\partial t^2} + \mathscr{C}_{s1}\frac{\partial^3 S_n}{\partial r^3}+ \mathscr{C}_{s2}\frac{\partial^3 S_n}{\partial t\partial r^2}\mathscr{C}_{s3}\frac{\partial^2 S_n}{\partial r^2}+\\
            &\mathscr{C}_{s4}\frac{\partial^2 S_n}{\partial t\partial r} + \mathscr{C}_{s5}\frac{\partial S_n}{\partial r} + \mathscr{C}_{s6}\frac{\partial S_n}{\partial t}  + \mathscr{C}_{s7} S_n = 0,
        \end{aligned}
        \label{eq:S_n_bulk}
    \end{equation}
\end{subequations}
where coefficients $\mathscr{C}_{(\cdot)}$, functions $\mathscr{G}_{(\cdot)}$ are in \cref{sec:appendix_coeff}.
Using the spherical harmonic projection of \cref{eq:tang_stress_continuity_BC} the tangential stress continuity condition for $T_n(r,t)$ and $S_n(r,t)$:
\begin{subequations}
\begin{equation}
    \begin{aligned}
       &\left(\zeta-4\lambda^7\mu\dot{R} \right)\frac{T_n(r,t)}{R^2}+\lambda^7\frac{\mu}{R}\left(\left.\frac{\partial T_n}{\partial t}\right|_{R}+\dot{R}\left.\frac{\partial T_n}{\partial r}\right|_{R} \right)=\\
       &\frac{2(2n+1)}{n+1}\left(R^{n-3}\kappa_n\left( \zeta+(n-5)\lambda^7\mu\dot{{R}}\right)+R^{n-2}\lambda^7\mu\dot{\kappa}_n\right)+\\
       &\frac{2(n+2)}{(n+1)R}\left(\epsilon_n\left(\zeta-3\lambda^7\mu \dot{R} \right)+\lambda^7\mu R \dot{\epsilon}_n\right),
    \end{aligned}
    \label{eq:T_b_bc}
\end{equation}
\begin{equation}
    \begin{aligned}
        &R \Bigg(\left.\frac{\partial S_n}{\partial r}\right|_{R}\left(\zeta-6 \lambda ^7 \mu  \dot{R}\right)+\lambda ^8 \mu  R_o \left(\dot{R} \left.\frac{\partial^2 S_n}{\partial r^2}\right|_{R}+\left.\frac{\partial^2 S_n}{\partial r\partial t}\right|_{R}\right) - \\
        & 2 \lambda ^7 \mu  \left.\frac{\partial S_n}{\partial t}\right|_{R}\Bigg) -2 S_n(R,t) \left(\zeta-5 \lambda ^7 \mu  \dot{R}\right) = 0, 
    \end{aligned}
    \label{eq:S_BC}
\end{equation}    
\end{subequations}
where $\zeta = G R_o \left(2 \alpha  \lambda ^6-3 \alpha  \lambda ^4+\alpha +\lambda ^4\right)$ and $\lambda=R/R_o$ is the radial stretch ratio at the bubble wall.
The bulk equation and boundary conditions for $S_n(r,t)$, \cref{eq:S_n_bulk,eq:S_BC} do not have source terms, without a constrain for $S_n(r,t)$ it is set to zero \citep{Prosperetti1977}. 

Integrating the radial component of \cref{eq:mom_balance} from the bubble wall to infinity, applying the pressure boundary equations \cref{eq:norm_stress_continuity_BC,eq:stress_equil_infty}, and using spherical harmonic orthogonality yields linear ordinary integro-differential equations for the modes.
For the zeroth mode and modelling weak material compression with far field acoustic radiation \citep{Keller1980}, the Keller-Miksis equation for the mean radius evolution is obtained:
\begin{equation}
    \begin{aligned}
        \left(1-\frac{\dot{R}}{c}\right)R\ddot{R} + &\frac{3}{2}\left(1-\frac{\dot{R}}{3c}\right)\dot{R}^2 = \\ 
        &\frac{1}{\rho}\left(1+\frac{\dot{R}}{c}+\frac{R}{c}\frac{d}{dt}\right) \left(p_{\text{b}}-p_{\infty}-\frac{2\gamma}{R}+S \right),
    \end{aligned}
    \label{eq:keller_miksis}
\end{equation}
where $c$ is the speed of sound in the material surrounding the bubble and $S$ the stress integral.
For the quadratic Kelvin-Voigt model~\cite{YANG2020}, the stress integral is 
\begin{equation}
\begin{aligned}
    S=  \frac{(3 \alpha-1) G}{2}&\left[5-\lambda^{-4}-4\lambda^{-1}\right]\\
    &+2 \alpha G\left[\frac{27}{40}+\frac{1}{8}\lambda^{-8}+\frac{1}{5}\lambda^{-5}+\lambda^{-2}-2\lambda\right] -\frac{4 \mu \dot{R}}{R}.
    \end{aligned}
\end{equation}
The far-field pressure takes the form $p_{\infty} = p_{\rm atm}-\mathcal{P}_a\sin(2 \pi f_at)$ where $p_{\rm atm}$ is the atmospheric pressure.
$\mathcal{P}_a$ and $f_a$ are the pressure amplitude and frequency of the ultrasound waves, respectively.
When ultrasound is not active, $\mathcal{P}_a=0$.
Applying orthogonality to modes $n\geq2$ yields the $n$-th integro-differential equation for the perturbation amplitude evolution:
\begin{equation}
    \ddot{\epsilon}_n+\eta\dot{\epsilon}_n+\xi\epsilon_n = 
    \mathcal{H}\ddot{\kappa}_n + \mathcal{G}\dot{\kappa}_n + \mathcal{C}\kappa_n + f_T+f_{\mathcal{J}}+f_\mathcal{K},
    \label{eq:pert_evo}
\end{equation}
where damping and stiffness coefficients $\eta$ and $\xi$, respectively, are in the supplementary information of \cite{Remillard2026} and the right-hand-side terms in \cref{sec:appendix_coeff}.
The mean radial oscillations account for heat and mass transfer as described in \cite{Estrada2018}.
Heat and mass transfer and weak material compression are small compared to the constitutive and inertial effects and are not accounted for directly in the evolution of the perturbations.
Additionally, the evolution of the mean radius (\cref{eq:keller_miksis}) is independent of the shape modes and therefore \cref{eq:keller_miksis,eq:pert_evo} are one-way coupled.

\subsection{Linear radial oscillations}
When the bulk radial motion is negligible, i.e., 
\begin{equation*}
    R(t) \approx R_o(1+\Delta\lambda(t)),
\end{equation*}
where $\Delta\lambda(t)  = R/R_o-1 = \mathscr{O}(\varepsilon)$, then $\dot{R} = \mathscr{O}(\varepsilon)$.
Applying this result to \cref{eq:pert_evo,eq:T_bulk,eq:T_b_bc}, then $\mathscr{O}(\varepsilon^2)$ terms and higher are neglected.
Therefore, for linear radial oscillations, \cref{eq:T_bulk} simplifies to 
\begin{equation}
     \begin{aligned}
        &\frac{\partial^2 T_n}{\partial t^2} - \frac{\partial^2}{\partial r^2}\left(\frac{\mu}{\rho}\frac{\partial T_n}{\partial t}+\frac{G}{\rho} \, T_n\right) + \frac{n(n+1)}{r^2}\left(\frac{\mu}{\rho} \frac{\partial T_n}{\partial t}  +\frac{G}{\rho}T_n\right)  = 0.
        \label{eq:T_bulk_lin}
        \end{aligned}
\end{equation}
When $G = 0$, this equation is the time derivative of the equation governing the rotational part of the velocity from \cite{Prosperetti1977} (see their equation 17).

The respective boundary condition (\cref{eq:T_b_bc}) and, after significant manipulation (see \cite{supplementary_information}), the perturbation evolution equation (\cref{eq:pert_evo}) simplify to 
\begin{equation}
    \begin{aligned}
        &G\frac{T_n(R,t)}{R_o}+\frac{\mu}{R_o}\left.\frac{\partial T_n}{\partial t}\right|_{R} = \\
        &\frac{2}{n+1}\left((2 n+1) R_o^{n-2} \left(G \kappa_n +\mu  \dot{\kappa}_n \right)+(n+2)  \left(G \epsilon_n+\mu  \dot{\epsilon}_n\right)\right),
    \end{aligned}
    \label{eq:T_BC_lin}
\end{equation}
\begin{equation}
    \begin{aligned}
        &\ddot{\epsilon}_n+\eta\dot{\epsilon}_n+\xi\epsilon_n = 
        -\frac{2n(n+2)}{\rho R_o^2}(\mu\dot\epsilon_n+G\epsilon_n) + \\
        & \frac{2n(n+1)(n+2)R_o^{n-4}}{\rho}\int_{R_o}^{\infty}s^{-n}\left(G(s,t) + \mu\,\frac{\partial T_n(s,t)}{\partial t} \right) ds.
    \end{aligned}
    \label{eq:simp_rhs}
\end{equation}
Neglecting the late-time rotation~\citep{Prosperetti1977}, the latter equation becomes: 
\begin{equation}
    \ddot{\epsilon}_n+ \frac{2(n+2)(2n+1)\mu}{\rho R_o^2}\dot{\epsilon}_n + \frac{2(n+2)}{\rho R_o^2}\left[(2n+1)G+\frac{(n+1)(n-1)\gamma}{2 R_o}\right]\epsilon_n = 0,
    \label{eq:lamb_linear_osc}
\end{equation}
with the damping coefficient predicted by \cite{Lamb}.
Additionally, the elastic contribution to the stiffness coefficient increases from the pure potential solution \citep{Remillard2026}, and has the same $n$ dependence as the damping coefficient.

\subsection{Boundary layer approximation}
The boundary layer approximation can be applied to the non-analytical integrals of \cref{eq:simp_rhs}.
Unlike \cite{Hilgenfeldt1996}, only the linearised radial motion regime eliminates the $T_n(r,t)$ dependence.
Thus, the integral on the right-hand side of \cref{eq:simp_rhs} simplifies to:
\begin{equation}
    \begin{aligned}
    \frac{2n(n+1)(n+2)R_o^{n-4}}{\rho}&\int_{R_o}^{\infty}s^{-n}\left(G T_n(s,t)+\mu\,\frac{\partial T_n(s,t)}{\partial t} \right)ds\approx \\
    &\frac{4n(n+2)^2\delta}{(1+2\delta)\rho R_o^2}\left(\mu\dot{\epsilon}_n + G \epsilon_n\right),
    \label{eq:BL_approx}
\end{aligned}
\end{equation}
where, $\delta$ is the nondimensional boundary layer thickness.
The boundary layer approximation predicts decreased damping and stiffness coefficients since $\delta$ is strictly positive.
\Cref{eq:BL_approx} matches that of \cite{Hilgenfeldt1996} when their radial oscillations are linearised, the perturbation amplitude is rewritten as $a_n =\epsilon_n R$, and $G=0$.

Following \cite{Hilgenfeldt1996}, we construct a similar heuristic viscoelastic generalisation of the classical viscous penetration depth by taking the Pythagorean sum of the viscous length scale with an analogous elastic length scale,
\begin{equation}
    \delta = \min \left(\frac{1}{R_o}\sqrt{\frac{1}{\rho\omega_c}\sqrt{\mu^2+\left(\frac{G}{\omega_c} \right)^2}}, \,  \frac{1}{2n}\right), 
    \label{eq:viscoelastic_del_cut}
\end{equation}
where $\omega_c$ is a characteristic frequency typically taken to be the forcing frequency.
When $G=0$, the boundary layer expression from \cite{Hilgenfeldt1996} is recovered.
For $\mu=0$, the boundary layer approximation becomes
\begin{equation*}
    \delta = \min \left(\frac{1}{R_o\omega_c}\sqrt{\frac{G}{\rho}}, \frac{1}{2n}\right),
\end{equation*}
where the first argument is the shear wave speed divided by a characteristic speed associated with the forcing.

\section{Problem setup\label{sec:problem_setup}}
\subsection{System of equations and nondimensionalisation\label{sec:systemequations}}

\begin{table}
\centering
\begin{tabular}{llll}
    Dimensional quantity & Range & Non-dimensional number & Range \\[1em]
    Equilibrium radius, $R_0$ & 10-$\SI{E3}{\micro\meter}$ & -- & -- \\
    Density, $\rho$ & \SI{1048}{\kilogram\per\meter^3} & -- & -- \\
    Ambient pressure, $p_{\infty}$ & \SI{101.3}{\kilo\pascal} & -- & --\\
    Shear modulus, $G$ & $1$-$\SI{E6}{}$ \SI{}{\pascal} & Cauchy number, $\Ca$ & $\SI{E-1}{}$-$\SI{E5}{}$\\
    Viscosity, $\mu$ & $\SI{E-5}{}$-10 \SI{}{\pascal\second} & Reynolds number, $\Rey$ & $\SI{E-2}{}$-$\SI{E6}{}$ \\
    Surface tension, $\gamma$ & $40$-$72$ \SI{}{\milli\newton/\meter}& Weber number, $\We$ & $7$-$905$ \\
    -- & -- & Strain stiffening index, $\alpha$ & $\SI{E-4}{}$-$\SI{E2}{}$\\
    -- & -- & Ohnesorge number, $\Oh$ & $\SI{2.57E-5}{}$-$292$\\
    -- & -- & Elastocapillary number, $\Ec$ & $\SI{8.93E-5}{}$-$8930$\\
    -- & -- & Mode number, $n$ & $2$--$13$ \\
\end{tabular}
\caption{Ranges of parameter values used in the numerical simulations.}
\label{tab:siminputs}
\end{table}

The full model is \cref{eq:T_bulk,eq:T_b_bc,eq:pert_evo}.
The semi-infinite spatial, time-dependent domain is mapped to a finite and stationary domain~\citep{Barajas2017}:
\begin{equation}
    x = 1-\frac{2}{1+\frac{r/R-1}{L}},
\end{equation}
where $x \in [-1, 1)$ is the transformed independent variable and $L$ scales the mapping and is set to $L=5$ for all simulations.
The system of equations can be written as
\begin{equation}
    \frac{\partial\mathbf{q}_n}{\partial t} + \mathbf{A}\frac{\partial \mathbf{q}_n}{\partial x}+ \mathbf{B}\frac{\partial^2 \mathbf{q}_n}{\partial x^2}+ \mathbf{C}\frac{\partial^3 \mathbf{q}_n}{\partial x^3}  +\mathbf{H}\mathbf{q}_n= \mathbf{G}(\mathbf{q}_n),
    \label{eq:system}
\end{equation}
where the state vector is
\begin{equation}
    \mathbf{q}_n = [T_n(r,t), \mathcal{V}_n(r,t), \epsilon_n(t), \dot{\epsilon}_n(t)]^{\T},
    \label{eq:system_o_eqns}
\end{equation}
with $\mathcal{V}_n = \partial_t T_n(r,t)|_r$.
$\mathbf{A}$, $\mathbf{B}$, and $\mathbf{C}$ are the advection, diffusion, and dispersion matrices, respectively.
$\mathbf{G}$ is the source term vector which includes the spatial integration terms. 
Vectors and matrices in \cref{eq:system_o_eqns} are defined in \cref{sec:appendix_coeff}.
Since the temporal discretisation is at fixed $x$, the $T_n$ and $\mathcal{V}_n$ equations have mapped domain advection-like terms, i.e., $-\partial x(r,t)/\partial t|_r (\partial T_n/\partial x)$ and $-\partial x(r,t)/\partial t|_r (\partial \mathcal{V}_n/\partial x)$. 

\Cref{tab:siminputs} summarises ranges of relevant dimensional and non-dimensional numbers used as inputs to the simulations.
The equations are nondimensionalised using the equilibrium radius ($R_o$), density ($\rho$), and pressure ($p_{\infty}$), unless stated otherwise. 
Thus, the characteristic time is $t_c = R_o\sqrt{\rho_c/p_c}$ and velocity $v_c = R_o/t_c$.
The respective Reynolds, Cauchy, Weber, Ohnesorge, and Elastocapillary numbers are
\begin{equation}
    \Rey = \frac{R_o\sqrt{\rho p_{\infty}}}{\mu}, \quad \Ca = \frac{p_{\infty}}{G}, \quad \We = \frac{R_o p_{\infty}}{2\gamma},
    \quad \Oh = \frac{\sqrt{\We}}{\Rey}, \quad \Ec = \frac{\We}{\Ca}.
    \label{eq:non_dim_numb}
\end{equation}
The $1/2$ factor is included in Weber number definition for consistency with \citet{Estrada2018} and \citet{pIMR2025}.
$\Oh$ is linearly proportional to $\mu$ and $\Ec$ is linearly proportional to $G$.
The characteristic scales were chosen such that one non-dimensionalisation could be used for free, ultra-sound forced, and LIC bubble oscillations.

\subsection{Numerical method}
The full model is solved using second order-accurate centred finite differences and trapezoidal numerical integration weights for the spatial derivatives and integrals respectively.
The second order backward differentiation formula (BDF2) is used to integrate the equations in time \citep{Hairer1993SolvingODEsI}.
Unless stated otherwise, the spatial and temporal domain, $x$ and $t$, is uniformly discretised into $N=256$ equidistant points and $\SI{E4}{}$ time-steps respectively for converged results (\cref{sec:convergence}).
The boundary condition \cref{eq:T_b_bc} replaces the first row of the $\mathcal{V}_n$ equations.
To satisfy kinematic equilibrium in the far field, $T_n$ and $\mathcal{V}_n$ are set to zero at $x=1$.
The characteristic speed associated with the advection due to domain mapping is $u(x,t) = \partial x(r,t)/\partial t|_r$, then to preserve numerical stability, these terms are upwinded as using the approximation to the advection term from the Beam-Warming method \citep{LeVeque2007FiniteDifference}.
The toroidal field and its time derivative are initially zero.

\begin{table}
\centering
\setlength{\tabcolsep}{6pt}
\begin{adjustbox}{max width=\textwidth}
\begin{tabular}{@{}L{0.15\textwidth} L{0.18\textwidth} 
C{0.25\textwidth} C{0.42\textwidth}@{}}

    Model & Rotational assumptions & $\eta$ & $\xi$ \\[1em]
    \cite{Remillard2026} & Potential-based & $\etaIrrot$ & $\xiRemillard$ \\ \addlinespace[0.35em]
    \cite{Prosperetti1977,Lamb} & Approximately irrotational & $\etaApprox$ & $\xiPlesset$ \\ \addlinespace[0.35em]
    Present model & Approximately irrotational & $\etaApprox$ & $\xiCurrentApprox$ \\ \addlinespace[0.35em]
    \cite{Hilgenfeldt1996} & Boundary Layer &$\etaBL$ & $\xiPlesset$  \\ \addlinespace[0.35em]
    Present model & Boundary layer & $\etaBL$ & $\xiCurrentBL$ \\
\end{tabular}
\end{adjustbox}
\caption{Constant damping and stiffness coefficients for different linear models, i.e., $\dot R =0$ and accurate up to $\mathscr{O}(\varepsilon)$. $\blcorr =(2n+1)-(2n(n+2)\delta)/(1+2\delta)$.}
\label{tab:models}
\end{table}

\subsection{Relative difference between models}
The simplified models are compared to the full model.
\Cref{tab:models} tabulates the damping and stiffness coefficients for different models when the radial oscillations are linearised, including the present full model. 
We use the relative differences in linear damping and stiffness of the models.
When radial oscillations are linear, \Cref{eq:pert_evo} can be approximated as
\begin{equation}
    \ddot{\epsilon}_n+ \overline{\eta}\dot{\epsilon}_n +\overline{\xi}\epsilon_n = 0,
    \label{eq:effective_mod}
\end{equation}
where $\overline{\eta}$ and $\overline{\xi}$ are the full model effective linear damping and stiffness coefficients, respectively.
These coefficients are evaluated by fitting \cref{eq:effective_mod} to the full model using $lsqcurvefit$ in MATLAB.
Then, relative differences between the coefficients for the full and reduced models are
\begin{equation}
    \mathcal{E}_{\eta} = 1-\frac{\eta}{\overline{\eta}}, \qquad  \mathcal{E}_{\xi} = 1 - \frac{\xi}{\overline{\xi}}.
\end{equation}

\section{Results and discussion\label{sec:results}}

\subsection{Infinitesimally small radial oscillations\label{sec:small_rad_osc}}

\begin{figure}
    \centering
    \includegraphics[width=\linewidth]{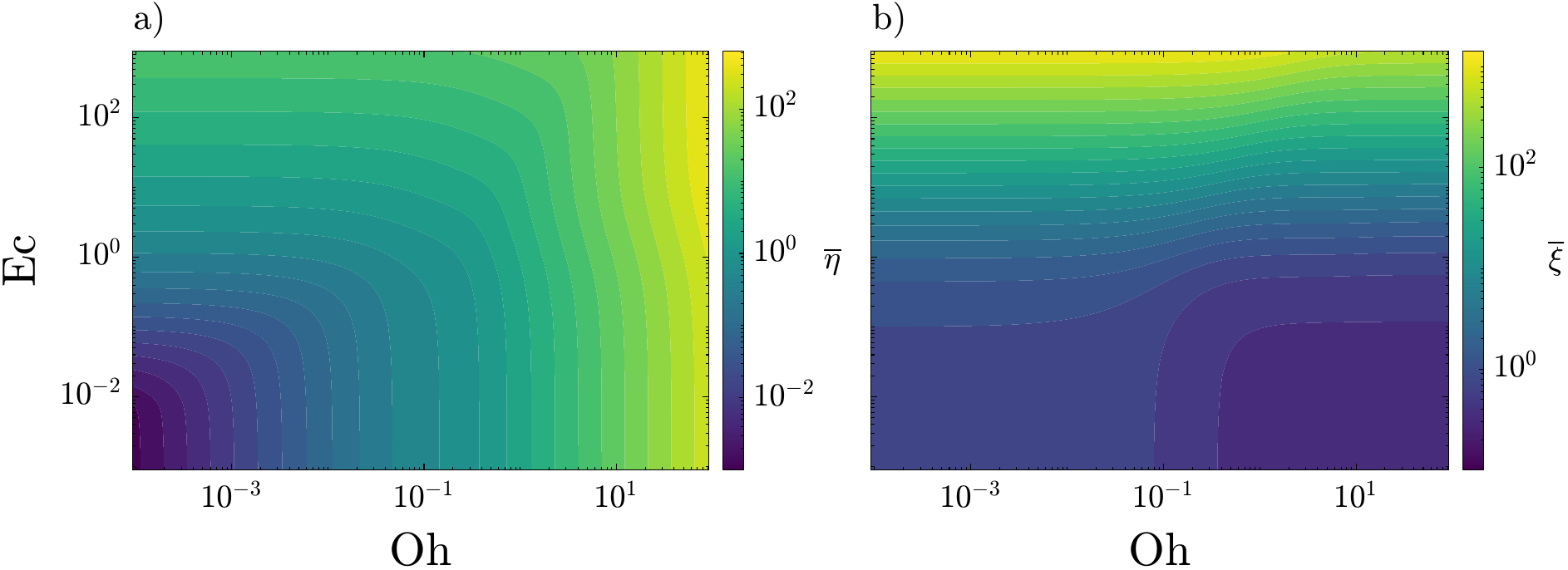}
    \caption{Contours of the effective full model damping (a) and stiffness (b) coefficients. 
    $R_o = \SI{100}{\micro\meter}$, $n=5$, $\We = 90.5$.}
    \label{fig:effective_coeffs}
\end{figure}
We first study the effective stiffness and damping coefficients of the full model in the limit of infinitesimally small radial oscillations. 
\Cref{fig:effective_coeffs} shows $\overline{\eta}$ and $\overline{\xi}$ as a function of $\Oh$ and $\Ec$ for a viscoelastic material surrounding the bubble.
Simulations were time-marched for $25$ periods using the non-dimensional natural frequency of oscillation predicted by the potential deformation model \cref{tab:models}.
As expected, $\overline{\xi}$ is approximately linearly proportional to $\Ec$ (see \cref{tab:models}).
Similarly, $\overline{\eta} \propto \Oh$ and $\overline{\eta} \propto \Ec$ thereby capturing an elastically driven dissipative mechanism.
On the other hand, the reduced models' damping coefficient only depends on $\Oh$.
The normalised $L^2$ difference of the perturbation amplitudes show the expected asymptotic behaviour as a function of $\Oh$ and agreement with \cite{Prosperetti1977} when elastic effects are neglected (see~\cref{sec:L2error}).

\begin{figure}
    \centering
    \includegraphics[width=\linewidth]{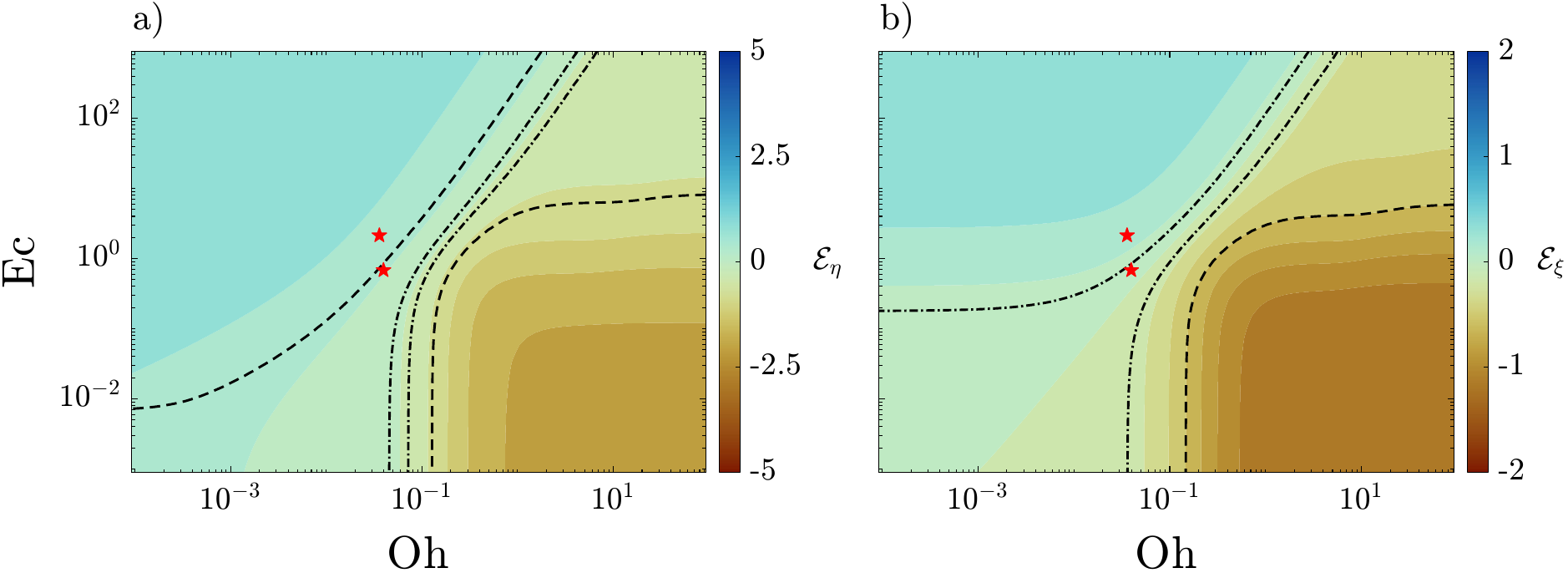}
    \caption{Contours of $\mathcal{E}_\eta$ (a), and $\mathcal{E}_\xi$ (b) for the potential-based model from \cref{tab:models}.
    Black dashed: $|\mathcal{E}_{(\cdot)}|= 0.5$, dashed-dotted lines: $|\mathcal{E}_{(\cdot)}|= 0.1$.
    Red stars: Averaged $\Ec$ and $\Oh$ values of ultrasound-driven experiments from \cite{Remillard2026}.
    $R_o = \SI{100}{\micro\meter}$, $n=5$, $\We = 90.5$.}
    \label{fig:potential_comp_VE}
\end{figure}

The potential-based model from \cref{tab:models} is compared to the full model to determine its regime of validity.
\Cref{fig:potential_comp_VE} shows the colour contours of $\mathcal{E}_\eta$ and $\mathcal{E}_\xi$ as functions of $\Oh$ and $\Ec$ for the potential-based model from \cref{tab:models}.
The results are independent of $\We$ and, thus, we set $n = 5$ and $\We =90.5$.
The small relative differences of the coefficients for their ultrasound experiments (see red stars) indicate that the potential-based modelling for their characterisation was appropriate.
As $\Ec \to 0$ and $\Oh \to 0$, the relative difference of the stiffness coefficient converges to zero, though the damping converges to a finite value.
Thus, for small viscoelastic effects, satisfying tangential stress continuity is necessary to accurately capture the damping coefficient.
Both the potential-based model and full model converge to the stiffness from \citet{Prosperetti1977} and \cite{Lamb} (see \cref{tab:models}).
On the other hand, the full model damping coefficient converges to that of Lamb and $\mathcal{E}_{\eta} \to -(2n+1)/(n+1)$. 
As $\Ec \to 0$ and $\Oh \to \infty$, the potential-based model from \cref{tab:models} over predicts the fitted coefficients.
The potential-based model stiffness coefficient discrepancy is due to the full model dependence on viscosity through the $\Oh$ number, while for the damping coefficient it is due to surrounding material rotations.
For the latter discrepancy, as viscosity increases, the potential-based model only increases tangential viscous traction.
On the other hand, the full model increases the rotational correction needed to counteract the potential contribution to the tangential viscous traction such that $\mathbf{n}\times(\boldsymbol{\sigma}_{\rm m}\mathbf{n}) =0$.
As $\Ec \to \infty$ and $\Oh \to 0$, the potential-based model under predicts both coefficients due to tangential stress continuity and presence of rotations.
Satisfying tangential stress continuity requires an increase in the rotational deformation field near the bubble surface. 
Consistent with the approximately irrotational model in \cref{tab:models}, the rotational deformation field contains strain energy that leads to a stiffened material response and larger stiffness for the full model.

\begin{figure}
    \centering
    \includegraphics[width=\linewidth]{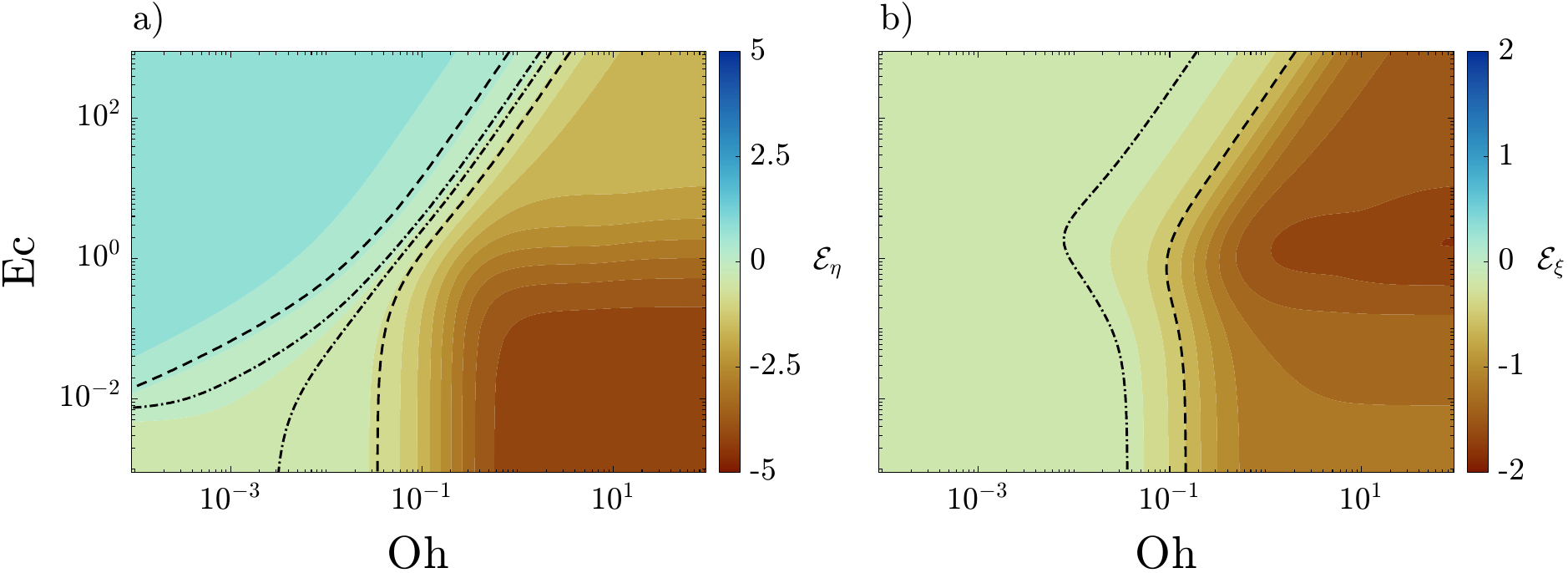}
    \caption{Contours of $\mathcal{E}_\eta$ (a), and $\mathcal{E}_\xi$ (b) for the approximately irrotational model \cref{eq:lamb_linear_osc}.
    Black dashed: $|\mathcal{E}_{(\cdot)}|= 0.5$; black dashed-dotted: $|\mathcal{E}_{(\cdot)}|= 0.1$.
    $R_o = \SI{100}{\micro\meter}$, $n=5$, and $\We =90.5$.
    }
    \label{fig:potential_comp_VE_lamb}
\end{figure}

\begin{figure}
    \centering
    \includegraphics[width=\linewidth]{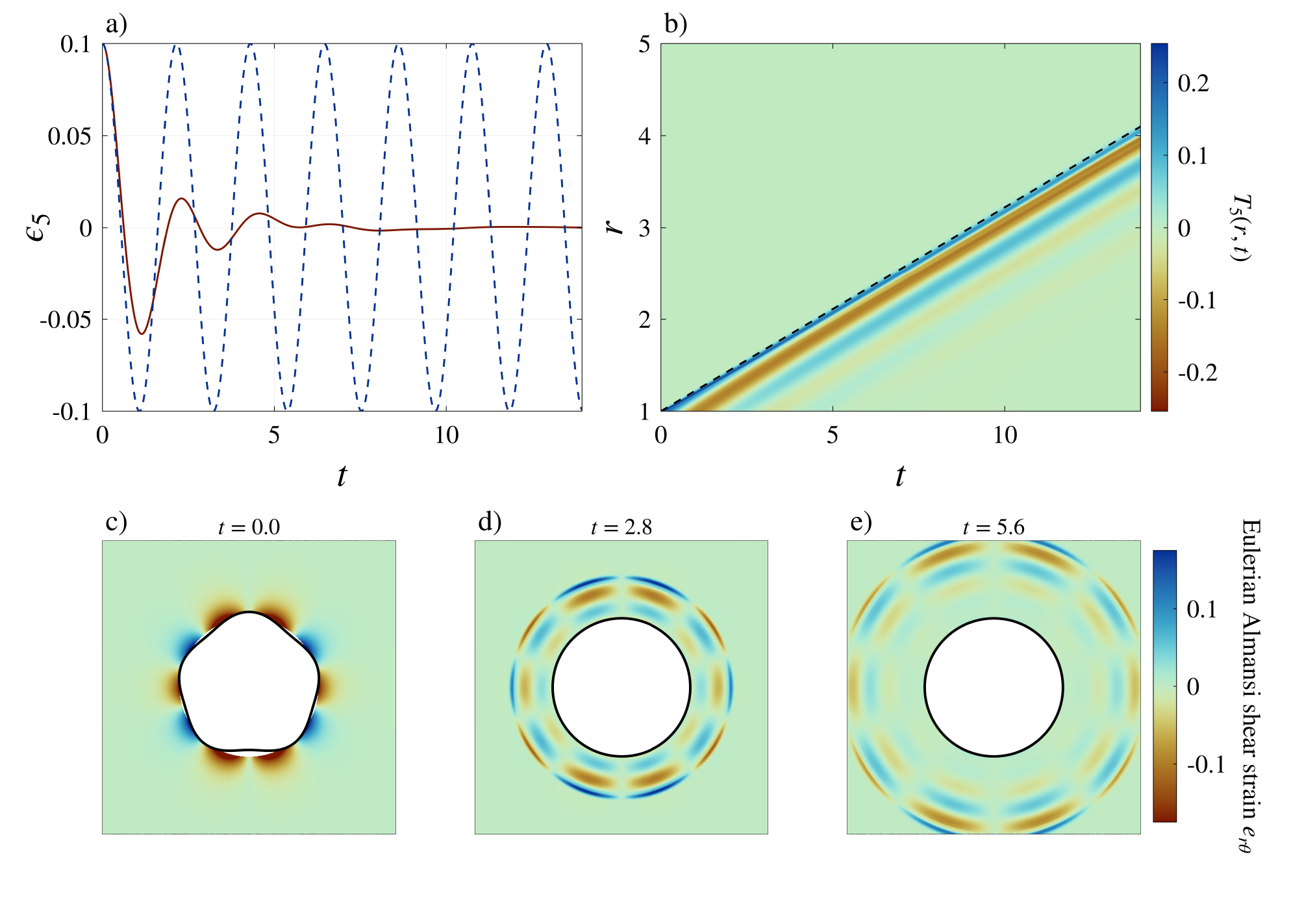}
    \caption{(a) Time history of a mode 5 surface perturbation in an elastic material. 
    Black solid line: full model solution, blue dashed line: approximately irrotational model \cref{eq:lamb_linear_osc}. 
    (b) The toroidal component of the deformation field.
    Black dashed line: $\sqrt{G/\rho}/v_c$.
    $R_o = \SI{100}{\micro\meter}$, $\Ec = 4.5$, $\We = 90.5$.
    (c)-(e) Contours of the Eulerian-Almansi shear strain, $e_{r\theta}$, for axisymmetric mode $5$ surface perturbation from the full model.
    }
    \label{fig:resolved_elastic_sim}
\end{figure}

Relative to the potential-based model, the approximately irrotational model (i.e., \cref{eq:lamb_linear_osc}) exhibits enhanced perturbation damping.
\Cref{fig:potential_comp_VE_lamb} shows $\mathcal{E}_{\eta}$ and $\mathcal{E}_{\xi}$ for the approximately irrotational model.
When viscous effects are small (i.e., $\Oh \ll \SI{E-1}{}$), inclusion of the infinitesimally thin layer confining the rotations approximates the full model linear stiffness coefficient.
Additionally, the asymptotic approximation is well suited for prediction of damping when both elastic and viscous effects are small, i.e., $\ll \SI{E-2}{}$.
Indeed, this model produces a larger area with the relative differences below $10\%$ than the potential-based model in  \cref{fig:potential_comp_VE}.
Including the boundary layer approximation (see \cref{eq:BL_approx} with $\omega_c = 2\pi/t_c$) provides marginal improvement to the reduced model accuracy (see \cref{sec:BL_approx}).
Reduced models inaccurately capture the $\epsilon_n$ evolution when elastic effects dominate (i.e., $\Ec > \SI{E-1}{}$ and $\Oh<\SI{E-3}{}\sqrt{\Ec}$) or viscous effects dominate ($\Oh>\SI{E-1}{}$ and $\Oh>\SI{E-3}{}\sqrt{\Ec}$).

\Cref{fig:resolved_elastic_sim} shows the mode $5$ time history (a) and the contour of the toroidal component of the deformation field in the elastically dominated regime (b).
The simulation is spatially resolved ($4096$ spatial points).
\Cref{fig:resolved_elastic_sim} also shows the corresponding colour contours of the full model shear strain $e_{r \theta}$ evolution (frame c to e):
\begin{equation*}
    e_{r \theta} = \frac{1}{r}\left[ \epsilon_n\frac{n+2}{n+1}\frac{R^{n+3}}{r^{n+2}}+\frac{1}{2}T_n(r,t)-\frac{\partial \Phi_n (r,t)}{\partial r} +\frac{\Phi_n(r,t)}{r}\right]\frac{\partial \YY}{\partial \theta }.
\end{equation*}
The oscillation frequencies match between the models but not the damping (frame a).
Since the potential-based model only admits purely monotonically decaying solutions, it localises the nonspherical strain energy near the bubble surface and decays spatially more rapidly than the full model.
On the other hand, for the full model, the majority of the torodial field has been emitted from the bubble surface by $t = 5$ as shear waves at speed $\sqrt{G/\rho}$ (see \cref{eq:T_bulk_lin}).
Due to tangential stress continuity and rotations, the strain energy delocalises from the surface, radiates outward, and drastically damps the surface perturbations.
As expected, the ten shear wave packets in the $\theta$ direction is twice the mode number, i.e., $\partial\YY/\partial \theta$ for $n =5$ has five local extrema from $\theta = [0, \pi]$.
Similarly, shear waves have been observed experimentally for nonspherical bubble oscillations in elastic gels \citep{Rapet2019}.

\subsection{Finite radial oscillations}
\subsubsection{Ultrasound-forced bubbles}
\begin{figure}
    \centering
    \includegraphics[width=\linewidth]{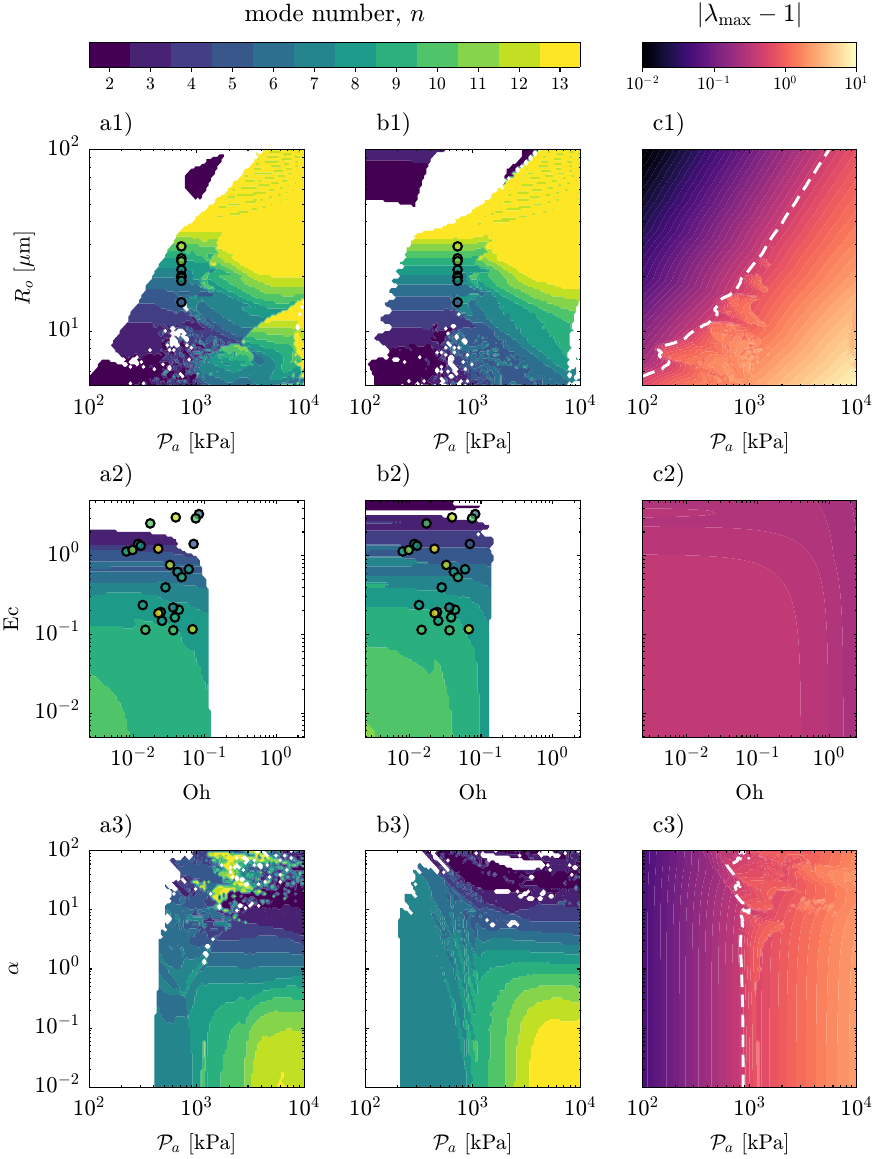}
    \caption{Ultrasound-forced bubble stability contour plots for modes $n = 2$--$13$ from the full model (a)  potential-based model (b) and the maximum stretch ratio at the bubble wall (c) versus $\Oh$ and $\Ec$ (row 1), $\mathcal{P}_a$ and $\alpha$ (row 2), and $\mathcal{P}_a$ and $R_o$ (row 3).  
    Colour circles: experimental data from \cite{Remillard2026}.
    White dashed line: $\left|\lambda_{\rm max}-1\right|=0.5$ and approximate transition region between parametric and Rayleigh-Taylor instabilities.
    $\gamma = \SI{0.04}{\newton\per\meter}$, $f_a = \SI{750}{\kilo\hertz}$.}
    \label{fig:para_inst}
\end{figure}
We study the stability of ultrasound-forced microbubbles by coupling \cref{eq:keller_miksis} with~\cref{eq:system}.
Modes $n = 2$--$13$ are initialised with an amplitude of $\SI{E-3}{}$ and evolved through $50$ cycles of ultrasound forcing using $\SI{5e3}{}$ time steps.
The simulation baseline values are $\mu = \SI{1.5}{\milli\pascal\second}$, $G = \SI{7.81}{\kilo\pascal}$, $R_o = \SI{20}{\micro\meter}$, $\mathcal{P}_a = \SI{750}{\kilo\pascal}$, and $\alpha=0$.
Surface tension and ultrasound-driving frequency are fixed: $\gamma = \SI{40}{\milli\newton\per\meter}$ and $f_a = \SI{750}{\kilo\hertz}$, respectively.
\Cref{fig:para_inst} shows contours of the most unstable mode number for the full model (a) and potential-based model \cref{tab:models} (b) versus $\mathcal{P}_a$ and $R_o$ (row 1), $\Oh$ and $\Ec$ (row 2), and $\mathcal{P}_a$ and $\alpha$ (row 3).
Contours of the bubble wall maximum stretch ratio are also shown (column c) and computed using~\cref{eq:keller_miksis}.
The approximate transition from parametric instability to the RTI for increasing $\lambda_{\text{max}}$ is indicated by the dashed white lines.
The experimental data from \cite{Remillard2026} show the parametric instability, with the most unstable mode inversely proportional to $\Ec$ and linear to $R_o$  \citep{Murakami2020,Hao1999,Hilgenfeldt1996}, and agrees with both the potential-based and full models (row 1 and 2).

The stability contours show that elasticity stabilises the perturbations, larger pressure amplitudes and equilibrium radii lead to higher unstable modes, and an instability-type transition with higher accelerations.
Rotations seem to generally stabilise the model indicated by more white space in column a compared to column b, consistent with previous modelling in viscous fluids \citep{Prosperetti1977,Hao1999}.
In the limit of small radial oscillations, the maximum stretch ratio at the bubble wall is linearly proportional to $\mathcal{P}_a$ and inversely to $R_o^{-1}$ \citep{Murakami2021} (c1).
Then, as ultrasound pressure amplitude increases and equilibrium radius decreases, both models show transition from completely stable, to parametric, to RTI (row 1).
Consistent with \cite{Remillard2026}, as $\Ec$ increases (i.e., shear modulus increases), the most unstable mode number decreases and transition to the RTI is not observed (row 2).
For fixed ground state material properties (i.e., $\mu$ and $G$) and $\mathcal{P}_a>\SI{E3}{\kilo\pascal}$, the instabilities transition from parametric to the RTI due to higher accelerations of the interface (row 3).
Additionally, presence of either parametric or RTI is approximately independent of $\alpha$.
For $\mathcal{P}_a>\SI{3E3}{\kilo\pascal}$, the approximately linearly inverse relationship between the most unstable mode and $\alpha$ is expected since $\alpha$ does not affect small radial oscillations for parametric instabilities.
Similar to $\Ec$, increasing $\alpha$ increases the nonlinear quadratic component of the elastic strain energy density function and suppresses the RTI.
Thus, a single experimental setup for linear to nonlinear elasticity characterisation could take advantage of this scaling by using the method from \cite{Remillard2026}.
Then, the applied ultrasound pressure wave amplitude can be increased to accurately characterise $\alpha$ using the RTI most unstable mode.

\subsubsection{Laser-induced cavitation}
\begin{table}
    \centering
    \begin{tabular}{c c c c c}
         Mode $n$ & Model & $\alpha = 0.09$ & $\alpha = 0$ & $\Delta$NRMSE \\[1em]
         0 (radial) & - & 0.055 & 0.069 & +0.014 \\[0.5em]
         \multirow{2}{*}{3, 4, 6, 7, 9, 10, 12} 
         & Full & 0.92 & 0.68 & -0.24\\
         & Potential-based & 0.77 & 1.1 & +0.33\\
    \end{tabular}
    \caption{Normalised root mean squared error (NRMSE) of the first bubble collapse from LIC.}
    \label{tab:nrmse_lic}
\end{table}

\begin{figure}
    \includegraphics[width=\linewidth]{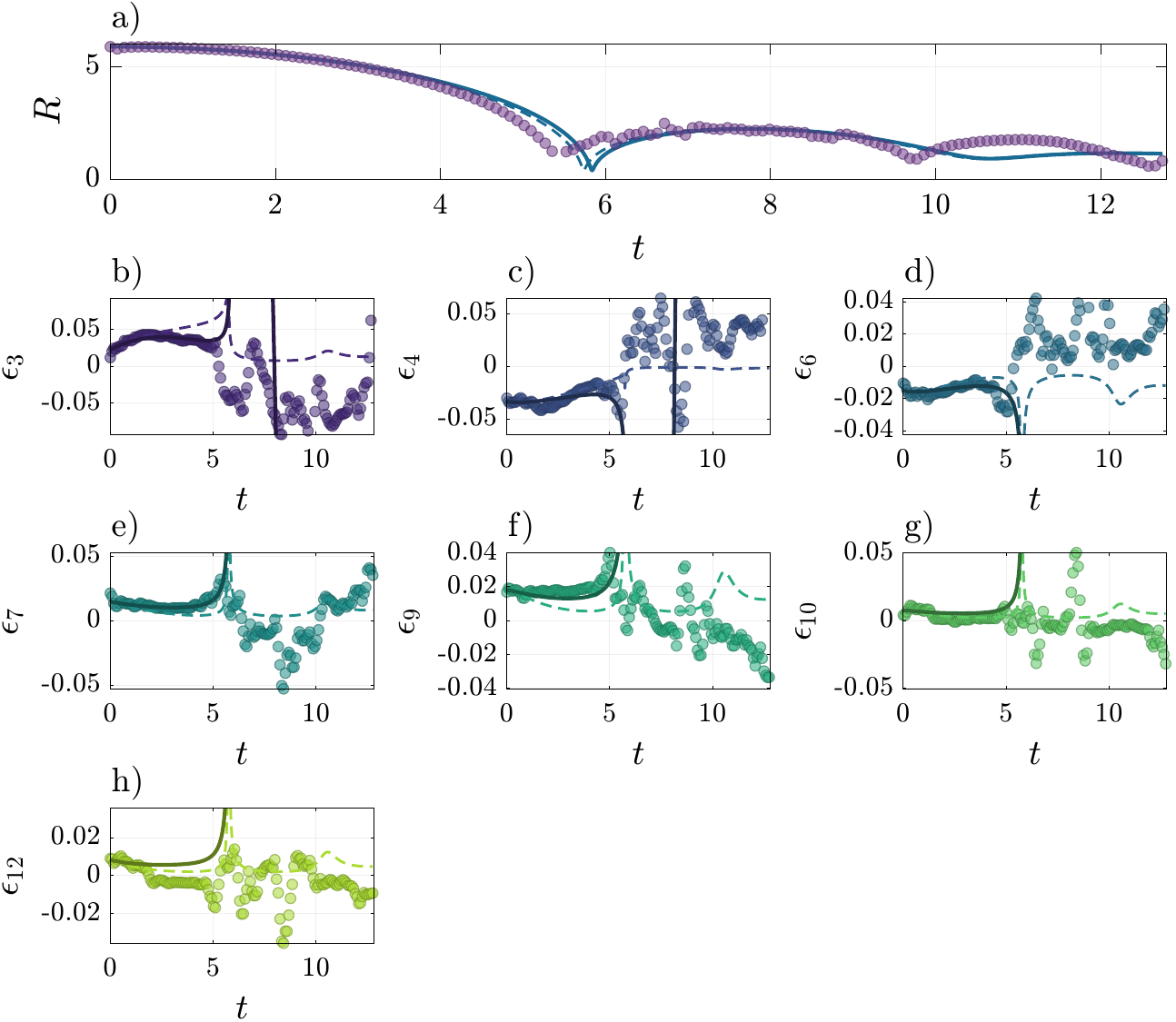}
    \caption{Mean radius (a) and shape modes (b)--(h) histories from the full model (solid lines, $\alpha = 0$), \cite{Remillard2026} (dashed lines, $\alpha = 0.090$), and LIC experimental data (circles) in polyacrylamide from \cite{YANG2020} and \cite{Remillard2026}. 
    $G =$ \SI{2.77}{\kilo\pascal} and $\mu =$ \SI{0.24}{\pascal\second}.}
    \label{fig:LIC_data_vs_model}
\end{figure}

To validate for inertial bubble collapse, we compare the LIC cavitation experiment data and potential-based model from \cite{Remillard2026} to the full model.
Their strain hardening parameter is derived by fitting with the experimental data and reported $\alpha=0.09$.
To minimise the mode oscillation frequency of the full model results and match the data, we also consider $\alpha = 0$.
\Cref{tab:nrmse_lic} tabulates the normalised root mean squared error (NRMSE) of the radial and shape mode data through the first bubble collapse for both models and $\alpha$ values and the NRMSE change.
Additionally, \Cref{fig:LIC_data_vs_model} shows the radial and mode histories from the experiments, the potential-based model with $\alpha = 0.09$, and the full model with $\alpha = 0$.
We focus on the first collapse since both linear models do not capture the subsequent nonlinear shape modes dynamics.
The increased discrepancy of the radial data for the full model with $\alpha = 0$ is small compared to the reduction in the NRMSE of the shape mode evolution. 
During the first collapse and without modelling the nonspherical modes, \citet{YANG2020} found $\alpha = 0.48$ matches the experimental radial data.
However, our improved shape mode fits with the full model indicate that the material may not be strain stiffening.
Instead, another strain or strain-rate-dependent mechanism, such as non-Newtonian viscosity or localised damage, may be taking place.

\begin{figure}
    \centering
    \includegraphics[width=\linewidth]{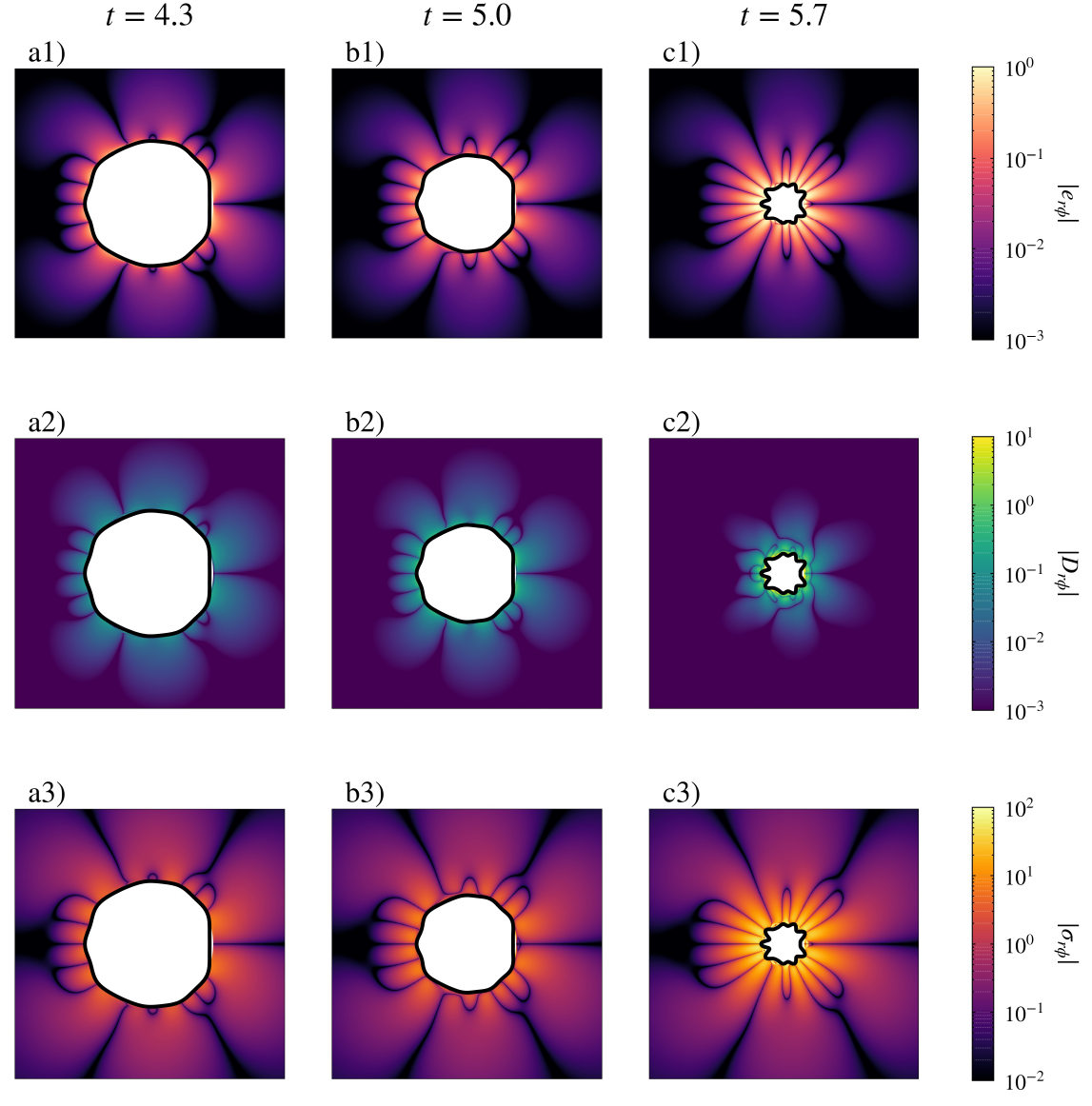}
    \caption{Contours of the Eulerian-Almansi shear strain $e_{r\phi}$ (row 1),  shear strain rate $D_{r\phi}$ (row 2), and shear stress $\sigma_{r\phi}$ (row 3) from the full model with contributions from modes $n=3$, $4$, $6$, $7$, $9$, $10$, and $12$. 
    Material properties \citep{Remillard2026}: $\mu = \SI{0.24}{\pascal \second}$, $G=\SI{2.77}{\kilo\pascal}$, $\gamma=\SI{56}{\milli\newton\per\meter}$, and $\alpha=0$.
    $\lambda_{\rm max} = 5.88$, $R_o=\SI{57.9}{\micro\meter}$, $\Ca = 36.6$, $\We = 52.2$, $\Rey=2.5$.
    }
    \label{fig:LIC}
\end{figure}
To gain insights into the dynamics during the first collapse, we consider the shear strains, strain-rates, and resulting stress response.
\Cref{fig:LIC} shows the full model Eulerian-Almansi shear strain $e_{r\phi}$,  shear strain rate $D_{r\phi}$, and shear stress $\sigma_{r\phi}$.
The respective dimensional shear strain rate and shear stress ranges are approximately $[\SI{1.7E3}{\per\second},\SI{1.7E6}{\per\second}]$ and $[\SI{1}{\kilo\pascal}, \SI{10}{\mega\pascal}]$.
During the majority of the bubble collapse, the perturbation amplitude is nearly constant.
Inertia, rather than asphericity-induced shear, dominates the dynamics in this early-time regime.
However, for the three quantities, a multi-petal pattern in the surrounding material intensifies as the bubble collapses, forming larger nonspherical perturbations at minimum bubble volume.
Comparing timescales viscous to elastic timescales yields $t_{\text{LIC}} = 0.915\lambda_{\text{max}}t_c$ yields $\mu/(Gt_{\text{LIC}}) = 2.7$.
Thus, the viscous resistance dominates early in the collapse and delays the strain and shear stress from returning toward its elastic equilibrium state and, thus, perturbation formation.
The larger six-petal pattern observed in the shear strain fields is a result of the dominant mode $3$, i.e., $|\partial\YY/\partial \phi|$ for $n = m = 3$ has six local maxima from $\phi = [0, 2\pi]$.
At $t = 5.7$, the higher mode numbers produce the largest strains near the bubble surface that decay radially and more rapidly than the lower modes.
Moreover, the maximum nonspherical shear and spherical radial normal strain rates are approximately $\textrm{max}(D_{r \phi}) \approx \SI{1.7E6}{\second^{-1}}$ and
\begin{equation*}
    \textrm{max}(D_{s,r r})  = 2\abs{\dot{R}/R} = \SI{1.3E6}{\second^{-1}},
\end{equation*}
respectively.
Since $\dot{R}$ is being underestimated at $t = 5.7$, these values are higher in the experimental observations.
For the stress response, \cite{YANG2020} predicts a maximum deviatoric radial normal elastic component of the stress of $\left|\sigma_{rr, \text{max}}^{e, \text{dev}}\right| = \SI{320}{\mega\pascal}$ at minimum volume.
However, by incorporating nonspherical rotations to more accurately capture the dynamics (see \cref{fig:LIC_data_vs_model}), we can reasonably assume that $\left| \sigma_{r\phi,\text{max}}\right|>\SI{10}{\mega\pascal}$ during collapse.
For the three quantities, the potential-based model predicts smaller values due to the potential deformation solution decaying as $\mathbf{u}_{\text{p}} \propto \nabla_{\mathbf{x}}(r^{-(n+1)})$ (see \cref{sec:potential_ss_predict}). 

The discrepancy between the experiment and models mean radius evolution is attributed to modelling limitations and underlying mechanisms that are difficult to access.
We assume axisymmetric bubble shape modes, i.e., $m = n$ \citep{Yang2021}; however, this might not be the case.
While earlier studies have shown that the shape modes for ultrasound experiments are axisymmetric \citep{Guédra_Inserra_2018,Philippe2018,Claude2017,Remillard2026}, there are few studies interrogating the axisymmetry of nonspherical bubbles under LIC loading \citep{Yang2021}.
Non-orthogonal contributions from out of plane shape modes may lead to a more rapid bubble collapse model prediction.
Similarly, the full model is linear and one-way coupled with energy going from the volume (radial) mode to the shape modes. 
Indeed, energy may transfer from the shape modes to the volume mode.
Concentrated and localised shear stresses from the aspherical collapse may disentangle the cross linkers leading to a time-dependent material response \cite{Spratt2021,Chu2025}.
These shear stresses could then induce plastic deformation, resembling the structures of defects imaged in similar agarose \citep{Yang2022} and PEGDA \citep{Luo2020} experiments.
Additional experimental, e.g., 3D tomography data \citep{Yang2022}, and numerical simulation modelling efforts are needed of nonspherical bubble collapse in soft viscoelastic materials to further understand this phenomenon.

\section{Conclusions\label{sec:conclusions}}

We derived a model for the evolution of the amplitude of shape modes of a gas bubble within a viscoelastic material with finite elasticity through a continuum mechanics framework.
Following \cite{Prosperetti1977}, a rotational deformation correction to the potential-based model of \cite{Remillard2026} was used to satisfy the traction conditions at the bubble wall and each component of the momentum balance.
The full model simplifies to that of \cite{Prosperetti1977} for linear radial oscillations and \cite{Lamb} when the rotations are confined to an infinitely thin layer surrounding the bubble to satisfy tangential stress continuity.
Accounting for the rotations, the perturbation evolution contains an elastic contribution to the stiffness coefficient with an equal mode number dependence to the damping coefficient from \cite{Lamb}.
When elastic effects dominate over viscous and capillary effects, radially outward propagating shear waves are emitted from the bubble surface. 
The parametric stability behaviour match between the full model, potential-based model, and experimental data for the conditions from \cite{Remillard2026}.
Under ultrasound forcing, the models disagree on the most unstable mode when the radial stretch ratio becomes large and the instabilities transition to RTI.
We find that the full model matches LIC experiments better when the strain-stiffening contributions are omitted, and the material may not be strain stiffening.
Discrepancies between the model and the experimental data may include material damage, plastic deformation, or nonlinear mode-mode and mode-volume coupling not included in this framework.
Future work will involve comparisons between the full model to 3D numerical simulations of nonspherical bubble collapse.

\backsection[Acknowledgements]{
The authors thank Profs. Jon Estrada and Jin Yang for the fruitful conversations during the preparation of this work.
OpenAI’s Codex was used to assist in coding and postprocessing, and all outputs were verified multiple times by the authors. 
Scientific content, reasoning, and conclusions are the authors' own.
}

\backsection[Funding]{
MRJ acknowledges support from the U.S. Department of Defense under the DEPSCoR program Award No. FA9550-23-1-0485 (PM Dr. Timothy Bentley) and U.S. National Science Foundation (NSF) under Grant No. 2232427.
This work used Anvil at Purdue University through allocation MCH220010 from the Advanced Cyberinfrastructure Coordination Ecosystem: Services \& Support (ACCESS) program, which is supported by U.S. National Science Foundation grants \#2138259, \#2138286, \#2138307, \#2137603, and \#2138296.
Funding agencies were not involved in study design; in the collection, analysis and interpretation of data; in the writing of the report; or in the decision to submit the article for publication.
The opinions, findings, and conclusions, or recommendations expressed are those of the authors and do not necessarily reflect the views of the funding agencies.} 

\backsection[Declaration of Interests]{The authors report no conflict of interest.}

\appendix
\section{Forward and inverse maps for coordinates\label{sec:appendix_maps}}

Using \cref{eq:pert_disp,eq:potential_displacments,eq:rot_disp,eq:map,eq:inv_map_eq}, the components of the forward $(r(\psi_o), \theta(\psi_o), \phi(\psi_o))$ and inverse maps $(r_o(\psi), \theta_o(\psi), \phi_o(\psi))$:

\begin{subequations}
\begin{equation}
    r(r_o, \theta_o, \phi_o, t) = r_s +  \Bigg[\epsilon_n\frac{R^{n+3}}{r_s^{n+2}}+\varepsilon T_n(r_s,t)-\varepsilon \left.\frac{\partial \Phi_n(r,t)}{\partial r}\right|_{r = r_s}\Bigg]Y_n^m(\theta_o, \phi_o), 
\end{equation}
\begin{equation}
\begin{aligned}
        \theta(r_o, \theta_o, \phi_o, t) &=\theta_o- \Bigg[\epsilon_n \frac{R^{n+3}}{(n+1)r_s^{n+3}}+\varepsilon\frac{\Phi_n(r_s,t)}{r_s^2}\Bigg]\frac{\partial}{\partial\theta_o}Y_n^m(\theta_o, \phi_o)\\
        &\qquad \, \, 
        + \varepsilon\frac{S_n(r_s,t)}{r_s^2}\csc\theta_o\frac{\partial}{\partial\phi_o}Y_n^m(\theta_o, \phi_o), 
\end{aligned}
\end{equation}
\begin{equation}
\begin{aligned}
    \phi(r_o, \theta_o, \phi_o, t) &= \phi_o -  \Bigg[\epsilon_n\frac{R^{n+3}}{(n+1)r_s^{n+3}}+\varepsilon \frac{\Phi_n(r_s,t)}{r_s^2}\Bigg]\csc^2\theta_o \frac{\partial}{\partial\phi_o}Y_n^m(\theta_o, \phi_o)\\
        &\qquad \, \, 
        - \varepsilon\frac{S_n(r_s,t)}{r_s^2}\csc\theta_o\frac{\partial}{\partial\theta_o}Y_n^m(\theta_o, \phi_o), 
\end{aligned}
\end{equation}    
\end{subequations}
\begin{subequations}
\begin{equation}
    r_o(r, \theta, \phi, t) = \varrho_s - \Bigg[\epsilon_n\frac{R^{n+3}}{r^n\varrho_s^{2}}+\varepsilon\frac{r^2}{\varrho_s^2}\left(T_n(r,t)-\frac{\partial\Phi_n(r,t)}{\partial r}\right)\Bigg]\YY,
\end{equation}
\begin{equation}
\begin{aligned}
    \theta_o(r, \theta, \phi, t) &=\theta +\Bigg[\epsilon_n \frac{R^{n+3}}{(n+1)r^{n+3}}+\varepsilon\frac{\Phi_n(r,t)}{r^2}\Bigg]\frac{\partial}{\partial\theta}\YY\\
    &\qquad \, \, 
        -  \varepsilon\frac{S_n(r,t)}{r^2}\csc\theta\frac{\partial}{\partial\phi}\YY,
\end{aligned}
\end{equation}
\begin{equation}
\begin{aligned}
        \phi_o(r, \theta, \phi, t) &= \phi + \Bigg[\epsilon_n \frac{R^{n+3}}{(n+1)r^{n+3}}+\varepsilon\frac{\Phi_n(r,t)}{r^2}\Bigg]\csc ^2\theta\frac{\partial }{\partial\phi}\YY\\
        &\qquad \, \, 
        +  \varepsilon\frac{S_n(r,t)}{r^2}\csc\theta\frac{\partial}{\partial\theta}\YY,
\end{aligned}
\end{equation}    
\end{subequations}
respectively.

\section{System of equations coefficients\label{sec:appendix_coeff}}
The expressions for $\eta$ and $\xi$ from \cref{eq:pert_evo} are in  \citet{Remillard2026}.
The forcing functions for \cref{eq:pert_evo}:
\begin{subequations}
\begin{equation}
\begin{aligned}
    f_T(t) &= \frac{-n(n+1)}{(n+4)(2n+1)R^3} \Bigg( n (n+1) \dot{R}^2 \,  T_n(R,t)-(n+4) R \Bigg(\ddot{R} T_n(R,t)\\
    &+\dot{R} \left(\dot{R} \left. \frac{\partial T_n}{\partial r}\right|_R +2 \left. \frac{\partial T_n}{\partial t}\right|_R \right)\Bigg) \Bigg) +\int_R^{\infty}\Bigg[n(n+1)^2\Big(\frac{R^2\dot{R}^2}{r^6}+\frac{1}{r^{10} \rho R^2}\\
    & \times\Big( -3 \alpha  G \varrho_s ^8+(3 \alpha -1) G \varrho_s ^4 r^4+4 \mu  r^5 R^2 \dot{R}\Big)\Big)T_n(r,t) \\
    &-n(n+1)^2\frac{\mu}{\rho r^2}\left( \frac{1}{R^2}\frac{\partial T_n}{\partial t}+\frac{\dot{R}}{r^2}\frac{\partial T_n}{\partial r}\right)\Bigg]dr,
    \end{aligned}
\end{equation}
\begin{equation}
    \begin{aligned}
        f_{\mathcal{J}}(t) &= \frac{n(n+1)^2}{(2n+1)} \int_R^\infty\Bigg[\frac{1}{r^{n+11}}\Bigg(-\frac{r^{9}}{R^2}\mathcal{J}_{l2}(r,t)+2 (n+2) r^6 \dot{R} \mathcal{J}_l(r,t) \\
        &+\mathcal{J}_n(r,t) \Bigg(\frac{nr^3}{R}\left( \dot{R}^2 \left(2 r^3-(n+7) R^3\right)+r^3 R \ddot{R}\right) + 6   n (n+2) r^4 \dot{R}\frac{\mu}{\rho} \\
        & +\frac{R_o}{\lambda^2 \varrho_s^7r}\frac{G}{\rho}(\lambda-1)(\lambda^2+\lambda+1)\Big(\alpha  \Big(3 \varsigma ^4 n (n+11)\\
        &+2 (n+2) (5 n+21) r^{12}-\varsigma  (27 n (n+7)+172) r^9+\varsigma ^2 (n (29 n+235)\\
        &+152) r^6-3 \varsigma ^3 (n (5 n+47)+16) r^3\Big)-(3 \alpha -1) \varrho_s ^2 r^4 \Big(\varsigma ^2 n (n+7)\\
        &+2 (n+2) (n+3) r^6-\varsigma  (n (3 n+17)+8) r^3\Big) \Big)\Bigg)\Bigg]dr,
    \end{aligned}
\end{equation}
\begin{equation}
    \begin{aligned}
        f_{\mathcal{K}} &= \int_R^\infty\frac{n(n+1)^2}{(2n+1)r^{-n+7}R^2}\Bigg( (n+1) R \, \mathcal{K}_n(r,t)\Big(\dot{R}^2 \left((n-6) R^3+2 r^3\right)\\
        &+r^3 R \ddot{R}\Big)+2 (n-1) r^3 R^2 \dot{R} \frac{\partial \mathcal{K}_n}{\partial t}+r^6 \frac{\partial^2 \mathcal{K}_n}{\partial t^2}+ \frac{1}{\rho r^4}\Bigg[\frac{G \varsigma}{\varrho_s^7} \Bigg(\alpha  \Big(3 \varsigma ^4 (n-10) \\
        & \times(n+1)+2 (n-1) (5 n-16) r^{12}+\varsigma  (-27 (n-5) n-10) r^9\\
        &+\varsigma ^2 (n (29 n-177)-54) r^6+3 \varsigma ^3 (n (37-5 n)+26) r^3\Big)\\
        &-(3 \alpha -1) \varrho_s ^2 r^4 \Big(\varsigma ^2 (n-6) (n+1)+2 (n-2) (n-1) r^6\\
        &+\varsigma  (n (11-3 n)+6) r^3\Big)\Bigg)+ 6 R^2 \mu  \left(n^2-1\right) r^5 \dot{R}\Bigg]\mathcal{K}_n(r,t)\Bigg)dr,
    \end{aligned}
\end{equation}
    \begin{equation}
    \begin{aligned}
        \mathcal{C} &=\frac{-n(4+n(n+8))}{n+4}R^{n-4}\left(R\ddot{R}+(n-1)\dot{R}^2 \right) \\
        &+\frac{n (n+1) \lambda ^{n-13} R_o^{n-5}}{\rho(n+6) (n+9) (n+12)}\\
        & \times\Bigg( G \left(\lambda ^3-1\right) R_o\Bigg[(3 \alpha -1) \lambda ^3 (n+13) \Big(2 \lambda ^6 (n+6) (n+9)\\
        &+\lambda ^3 n (n (3 n+35)+86)-n (n+2) (n+7)\Big)\\
        & \times \, _2F_1\left[(-\frac{1}{3},\frac{n}{3}+4;\frac{n}{3}+5;1-\frac{1}{\lambda ^3}\right]+(3 \alpha -1) \Big(-2 \lambda ^9 (n+6) \\
        &\times(n+9)+\lambda ^6 n (n+1) (n+6) (n+9)-\lambda ^3 n (n+2) (n+7) (2 n+19)\\
        &+n (n+2)(n+7) (n+10)\Big) \, _2F_1\left[\frac{2}{3},\frac{n}{3}+4;\frac{n}{3}+5;1-\frac{1}{\lambda ^3}\right]\\
        &+2 \alpha  \Big(4 \lambda ^{12} (n+6) (n+9)+2 \lambda ^9 (n+1) (n+2) (n+6) (n+9)\\
        &+\lambda ^6 (n+1) (n+6) (n+9) (n (n+6)+6)\\
        &-\lambda ^3 n (n (n (2 n (n+25)+427)+1411)+1398)+n (n+2) (n+11) \\
        &\times(n (n+13)+39)\Big) \, _2F_1\left[\frac{1}{3},\frac{n}{3}+4;\frac{n}{3}+5;1-\frac{1}{\lambda ^3}\right]-\frac{\alpha}{\lambda^2}  (n+14)\\
        & \times\Bigg(4 \lambda ^9 (n+6) (n+9)+2 \lambda ^6 (n+1) (n+2) (n+6) (n+9)\\
        &-\lambda ^3 (n (n (n (4 n+59)+229)+120)-324)+n (n (n (2 n+33)+175)\\
        &+318)\Bigg) \, _2F_1\left[1,\frac{n+17}{3};\frac{n}{3}+5;1-\frac{1}{\lambda ^3}\right]\Bigg]\Bigg)-\frac{2n(n+1)(2n+1)R^n}{\rho R_o^5\lambda^{12}}\\
        & \times \Bigg( G R_o \left(\alpha  \left(\lambda ^6-3 \lambda ^4+2\right)+\lambda ^4\right)+\lambda ^7 \mu  (n-2) \dot{R}\Bigg),
    \end{aligned}
    \end{equation}
    \begin{equation}
        \mathcal{G} = \frac{-2n(2+n(n+6))}{n+4}R^{n-3}\dot{R}-2n(n+1)(2n+1)R^{n-4}\frac{\mu}{\rho},
    \end{equation}
    \begin{equation}
        \mathcal{H} = -nR^{n-2},
    \end{equation}
\end{subequations}
where 
\begin{subequations}
    \begin{equation*}
        \varsigma = R_o^3(\lambda^3-1),
    \end{equation*}
    \begin{equation*}
        \mathcal{J}_l(r,t) = \int_R^rs^{n+1}\frac{\partial T_n}{\partial t}ds,
    \end{equation*}
    \begin{equation*}
        \mathcal{J}_{l2}(r,t) = \int_R^rs^{n+1}\frac{\partial^2 T_n}{\partial t^2}ds,
    \end{equation*}
\end{subequations} and $_2F_1$ is the Gauss hypergeometric function.

The coefficients for the toroidal field PDE are, 
\begin{subequations}
    \begin{equation}
            \mathscr{C}_{t1} = - \frac{R^2\dot{R}}{r^2}\frac{\mu}{\rho},
    \end{equation}
    \begin{equation}
        \mathscr{C}_{t2} = -\frac{\mu}{\rho},
    \end{equation}
    \begin{equation}
        \mathscr{C}_{t3} = \frac{R^4\dot{R}^2}{r^4}-\frac{G}{\rho} \frac{\varrho_s^2}{r^8} \left(2 \alpha  \varsigma ^2+4 \alpha  \varsigma  \varrho_s ^3+3 \alpha  \varrho_s ^6+(1-3 \alpha ) \varrho_s ^2 r^4 \right)+8 \frac{\mu}{\rho}   \frac{R^2 \dot{R}}{r^3}, 
    \end{equation}
    \begin{equation}
        \mathscr{C}_{t4} = 2\frac{R^2\dot{R}}{r^2},
    \end{equation}
    \begin{equation}
    \begin{aligned}
        \mathscr{C}_{t5} &= \frac{R}{r^5}\left(r^3R \ddot{R}+2 \left(r^3-4R^3\right)\dot{R}^2\right) - \frac{8\varsigma}{\varrho_s \, r^{9}}  \Big(\alpha  \varsigma ^2+2 \alpha  \varsigma  \varrho_s ^3+3 \alpha  \varrho_s ^6\\
        &+(1-3 \alpha ) \varrho_s ^2 r^4\Big)\frac{G}{\rho} +\frac{\left(n^2+n-36 \right)R^2\dot{R}}{r^4}\frac{\mu}{\rho},
    \end{aligned}
    \end{equation}
    \begin{equation}
        \mathscr{C}_{t6} = -\frac{6 R^2 \dot{R}}{r^3} +\frac{n(n+1)}{r^2}\frac{\mu}{\rho},
    \end{equation}
    \begin{equation}
    \begin{aligned}
        \mathscr{C}_{t7} &= -\frac{2 R}{r^6}\left( 2 r^3 R \ddot{R}+\left(4 r^3-13 R^3\right) \dot{R}^2\right) + \frac{1}{r^{10} \varrho_s^4}\frac{G}{\rho}\Big((1-3 \alpha ) \varrho_s ^2 r^4 \Big(\varsigma^2 \\
        &\times\left(n^2+n-6\right)+2 \varsigma \left(n^2+n+10\right) \varrho_s ^3+n (n+1) \varrho_s ^6\Big)+\alpha  \Big(2 \varsigma^4 \\
        &\times\left(n^2+n-43\right)-8 \varsigma^3 \left(n^2+n-27\right) r^3+3 n (n+1) r^{12}-2 \varsigma (n-5) \\
        &\times (n+6) r^9+\varsigma^2 (9 n (n+1)-190) r^6\Big)\Big)+2\frac{n^2+n+48}{r^5}\frac{R^2\dot{R} \mu}{\rho},
        \end{aligned}
    \end{equation}
    \begin{equation}
    \begin{aligned}
        \mathscr{G}_{\Phi_n} &= \frac{6n(n+2)R^{2n+2}}{(n+1)r^{n+8}}\Bigg(\kappa_n (t) \Big(\dot{R}^2 \left((2 n+3) r^3-(n+7) R^3\right)+r^3 R \ddot{R}\Big)\\
        &+r^3 R \dot{R} \dot{\kappa}_l(t)\Bigg)+ \frac{6n(n+2)R}{(2n+1)r^{n+8}}\Bigg(\mathcal{J}_n(r,t) \Big(\dot{R}^2 \left(2 r^3-(n+7) R^3\right)\\
        &+r^3 R \ddot{R}\Big)+r^3 R \dot{R} \mathcal{J}_l(r,t) \Bigg) -\frac{6(n^2-1)R}{(2n+1)r^{-n+7}}\Bigg(\mathcal{K}_n(r,t) \Big(\dot{R}^2 \Big((n-6) \\
        &\times R^3+2 r^3\Big)+r^3 R \ddot{R}\Big)+r^3 R \dot{R} \frac{\partial \mathcal{K}_n}{\partial t}\Bigg)- \Big(\kappa_n(t) +\frac{(n+1)}{(2n+1)R^{2n+1}}\\
        &\times \mathcal{J}_n(r,t)\Big)\frac{2 n(n+2)  R^{2 n+1}}{(n+1) \rho\varrho_s ^7r^{n+12}}\Bigg(G \varsigma \Big(-\alpha  \left(n^2+n+42\right) \varrho_s ^9 r^3\\
        &-\alpha  (n-6) (n+11) \varrho_s ^{12}+8 \alpha  n r^{12}-(3 \alpha -1) (n-1) \varrho_s ^2 r^{10}\\
        &+\alpha  (n+1) (n+3) \varrho_s ^3 r^9-(3 \alpha -1) (n-5) \varrho_s ^5 r^7+\alpha  (n+3) (n+6) \\
        & \times\varrho_s ^6 r^6-3 (3 \alpha -1) (n+7) \varrho_s ^8 r^4\Big)+30 \mu  (n+3) \varrho_s ^7 r^5 R^2 \dot{R}\Bigg) \\
        &+\frac{2 (n^2-1)\mathcal{K}_n(r,t)}{(2 n+1) \rho\varrho_s ^7 \, r^{-n+11}}\Bigg(G \varsigma \Big(\alpha\Big(\left(-n^2-n-42\right)\varrho_s ^9 r^3-(n-10) \\
        & \times(n+7) \varrho_s ^{12}-8 (n+1) r^{12}+(n-2) n \varrho_s ^3 r^9\\
        &+(n-5) (n-2) \varrho_s ^6 r^6\Big)+(3 \alpha -1) \varrho_s ^2 r^4 \Big(3 (n-6) \varrho_s ^6+(n+2) r^6\\
        &+(n+6) \varrho_s ^3 r^3\Big)\Big)-30 \mu  (n-2) \varrho_s ^7 r^5 R^2 \dot{R}\Bigg),
    \end{aligned}
    \end{equation}
    \begin{equation}
    \begin{aligned}
        \mathscr{G}_{\rm p} &= 6\frac{n+2}{n+1}\frac{R^{n+4}}{r^{n+8}}\Bigg( \epsilon_n \left(\dot{R}^2 \left((n+5) r^3-(n+7)R^3\right)+r^3R \ddot{R}\right)\\
        &+r^3R\dot{R} \dot{\epsilon}_n\Bigg)  -2\epsilon_n\frac{(n+2)  }{(n+1) }\frac{R^{n+3}}{\rho  \varrho_s^{7}r^{n+12}}\Bigg( G \varsigma\Big(-\alpha  \left(n^2+n+42\right) \varrho_s ^9 r^3\\
        &-\alpha  (n-6) (n+11) \varrho_s ^{12}+8 \alpha  n r^{12}-(3 \alpha -1) (n-1) \varrho_s ^2 r^{10}+\alpha  (n+1) \\
        & \times(n+3) \varrho_s ^3 r^9-(3 \alpha -1) (n-5) \varrho_s ^5 r^7+\alpha  (n+3) (n+6) \varrho_s ^6 r^6\\
        &-3 (3 \alpha -1) (n+7) \varrho_s ^8 r^4\Big)+30 \mu  (n+3) \varrho_s ^7 r^5 R^2 \dot{R}\Bigg),
    \end{aligned}
    \end{equation}
    \label{eq:T_bulk_coeffs}
\end{subequations}
where, 
\begin{equation*}
     \dot{\kappa}_l(t) = -\frac{n+1}{2n+1}\int_{R}^{\infty} s^{-n} \frac{\partial T_n}{\partial t} ds.
\end{equation*}
and for the poloidal field, 
\begin{subequations}
    \begin{equation}
        \mathscr{C}_{s1} = - \frac{R^2\dot{R}}{r^2}\frac{\mu}{\rho},
    \end{equation}
    \begin{equation}
        \mathscr{C}_{s2} = -\frac{\mu}{\rho},
    \end{equation}
    \begin{equation}
    \begin{aligned}
        \mathscr{C}_{s3} &=  \frac{1}{\rho \, r^8}\left((3 \alpha -1) G \varrho_s ^4 r^4-\alpha  G \varrho_s ^2 \left(\varsigma ^2+3 r^6-2 \varsigma  r^3\right)+6 \mu  r^5 R^2 \dot{R} \right)\\
        &+\frac{R^4\dot{R}^2}{r^4},
    \end{aligned}
    \end{equation}
    \begin{equation}
        \mathscr{C}_{s4} = 2\frac{R^2\dot{R}}{r^2},
    \end{equation}
    \begin{equation}
    \begin{aligned}
        \mathscr{C}_{s5} &= \frac{R}{r^5}\left(2(r^3-2R^3)\dot{R}^2+r^3R\ddot{R} \right)-4\frac{G\varsigma}{\rho\varrho_s r^9}\Big( 2 \alpha  \varsigma ^2+3 \alpha  r^6\\
        &+(1-3 \alpha ) \varrho_s ^2 r^4-4 \alpha  \varsigma  r^3\Big)+(n^2+n-18)\frac{R^2\dot{R}\mu}{\rho \, r^4},
    \end{aligned}
    \end{equation}
    \begin{equation}
        \mathscr{C}_{s6} = \frac{n(n+1)}{r^2}\frac{\mu}{\rho}-2\frac{R^2\dot{R}}{r^3},
    \end{equation}
    \begin{equation}
    \begin{aligned}
        \mathscr{C}_{s7} &=-2\frac{R}{r^6}\left((2r^3-3R^3)\dot{R}^2+r^3R\ddot{R} \right)+\frac{G}{\rho r^{10}\varrho_s^4}\Big(\alpha   \Big(-14 \varsigma ^4+3 n \\
        & \times(n+1) r^{12}-2 \varsigma  (n-2) (n+3) r^9+\varsigma ^2 (n-6) (n+7) r^6+40 \varsigma ^3 r^3\Big)\\
        &-(3 \alpha -1) \varrho_s ^2 r^4 \left(-6 \varsigma ^2+n (n+1) r^6+4 \varsigma  r^3\right) \Big)\\
        &-2(n-3)(n+4)\frac{R^2\dot{R}\mu}{r^5\rho}.
           \end{aligned}
    \end{equation}
    \label{eq:S_bulk_coeffs}
\end{subequations}

The matrices that are described in \cref{eq:system} are 
\begin{subequations}
    \begin{equation}
    \mathbf{A} = \begin{bmatrix}u(x,t) & 0 & 0 & 0\\
    \mathscr{C}_{t1}\mathscr{X}_3^1 +\mathscr{C}_{t3}\mathscr{X}_2^1 +\mathscr{C}_{t5}\mathscr{X}_1^1 & u(x,t) + \mathscr{C}_{t2}\mathscr{X}_2^1 + \mathscr{C}_{t4}\mathscr{X}_1^1 & 0 & 0\\
    0 & 0 & 0 & 0\\
    0 & 0 & 0 & 0\\
    \end{bmatrix},
    \end{equation}
    \begin{equation}
    \mathbf{B} = \begin{bmatrix}0 & 0 & 0 & 0\\
    \mathscr{C}_{t1}\mathscr{X}_3^2 +\mathscr{C}_{t3}\mathscr{X}_2^2 & \mathscr{C}_{t2}\mathscr{X}_2^2 & 0 & 0\\
    0 & 0 & 0 & 0\\
    0 & 0 & 0 & 0\\
    \end{bmatrix},
    \end{equation}
    \begin{equation}
    \mathbf{C} = \begin{bmatrix}0 & 0 & 0 & 0\\
     \mathscr{C}_{t1} \mathscr{X}_3^3 & 0 & 0& 0\\
    0 & 0 & 0 & 0\\
    0 & 0 & 0 & 0\\
    \end{bmatrix},
    \end{equation}
    \begin{equation}
    \mathbf{G} = \begin{bmatrix} 0 \\
    -\mathscr{G}_{\Phi_n} -\mathscr{G}_{\rm p} \\
    0 \\
    \mathcal{H}\ddot{\kappa}_n + \mathcal{G}\dot{\kappa}_n + \mathcal{C}\kappa_n + f_T+f_{\mathcal{J}}+f_\mathcal{K}\\
    \end{bmatrix},
    \end{equation}
    \begin{equation}
    \mathbf{H} = \begin{bmatrix}0 & -1 & 0 & 0\\
    \mathscr{C}_{t7} & \mathscr{C}_{t_6} & 0 & 0\\
    0 & 0 & 0 & -1\\
    0 & 0 & \xi & \eta\\
    \end{bmatrix},
    \end{equation}
\end{subequations}
where
\begin{subequations}
    \begin{equation}
        \frac{\partial}{\partial r} = \frac{\partial x}{\partial r} \frac{\partial}{\partial x} = \frac{(1-x)^2}{2 R L}\frac{\partial}{\partial x} = \mathscr{X}_1^1\frac{\partial}{\partial x},
    \end{equation}
    \begin{equation}
    \begin{aligned}
        \frac{\partial^2}{\partial r^2} &= \left(\frac{\partial x}{\partial r} \right)^2\frac{\partial^2}{\partial x^2}+\frac{\partial x}{\partial r}\left[\frac{\partial}{\partial x}\frac{\partial x}{\partial r}
        \right]  \frac{\partial}{\partial x}
        \\& =  
        \frac{(1-x)^4}{4 R^2 L^2}\frac{\partial^2}{\partial x^2}-\frac{(1-x)^3}{2 R^2 L^2}\frac{\partial}{\partial x}=\mathscr{X}_2^2\frac{\partial^2}{\partial x^2}-\mathscr{X}_1^2\frac{\partial}{\partial x},
    \end{aligned}        
    \end{equation}
\end{subequations}
\begin{equation}
    \begin{aligned}
        \frac{\partial^3}{\partial r^3} &= \left(\frac{\partial x}{\partial r}\right)^3\frac{\partial^3}{\partial x^3}+3\left(\frac{\partial x}{\partial r}\right)^2\left[\frac{\partial}{\partial x}\frac{\partial x}{\partial r}\right]\frac{\partial^2}{\partial x^2}+\Bigg[\left(\frac{\partial x}{\partial r}\right)^2\left[\frac{\partial^2}{\partial x^2} \frac{\partial x}{\partial r}\right]\\
        &+\frac{\partial x}{\partial r}\left[\frac{\partial}{\partial x}\frac{\partial x}{\partial r}\right]^2\Bigg]\frac{\partial}{\partial x}\\
        & = \frac{(1-x)^6}{8 R^3 L^3}\frac{\partial^3}{\partial x^3}-3\frac{(1-x)^5}{4 R^3 L^3}\frac{\partial^2}{\partial x^2}+3\frac{(1-x)^4}{4 R^3 L^3}\frac{\partial }{\partial x}\\
        &=\mathscr{X}_3^3\frac{\partial^3}{\partial x^3}+\mathscr{X}_3^2\frac{\partial^2}{\partial x^2}+\mathscr{X}_3^1\frac{\partial }{\partial x}.
        \end{aligned}
\end{equation}

\section{Convergence\label{sec:convergence}}
\begin{figure}
    \centering
    \includegraphics[width=0.75\linewidth]{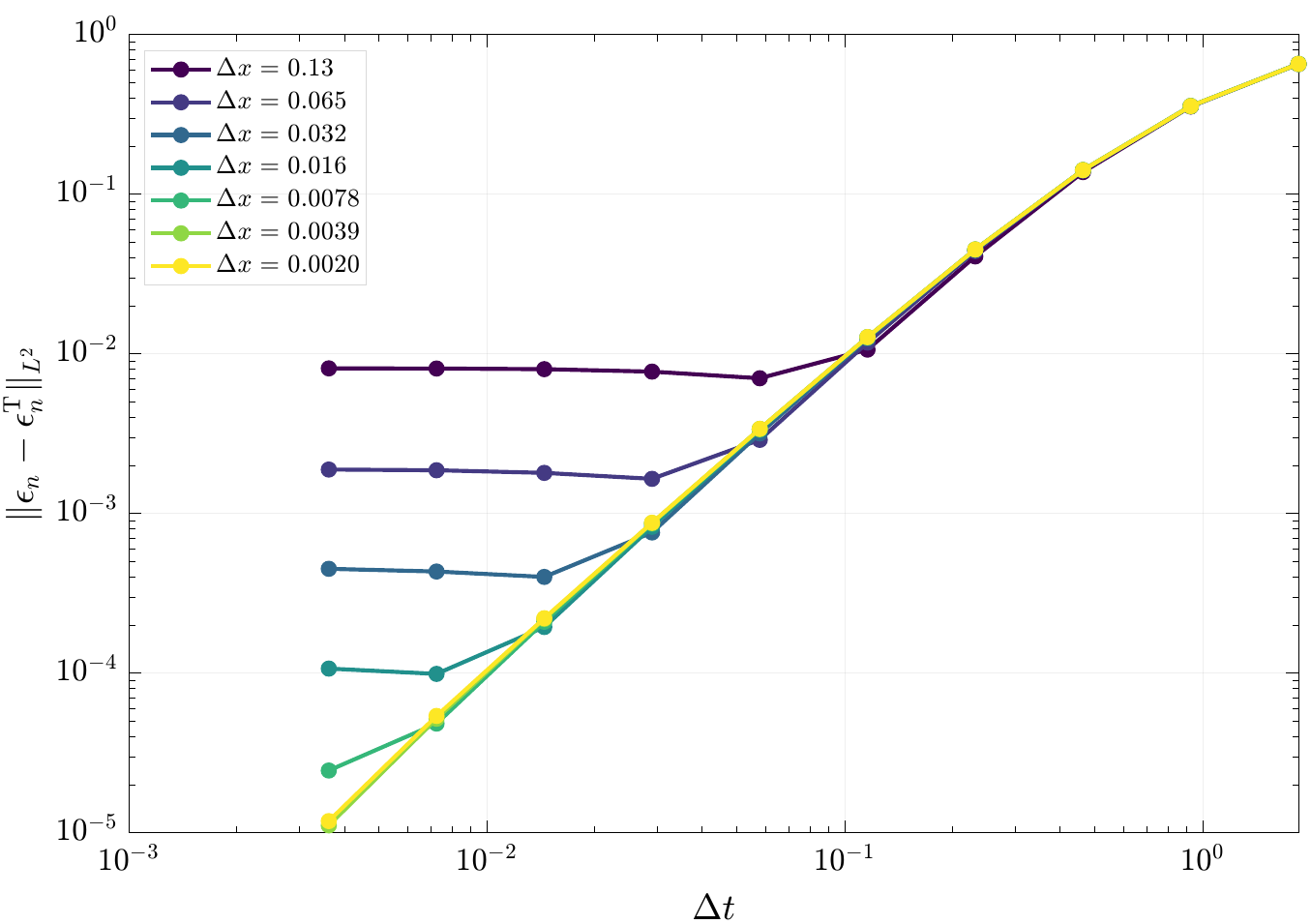}
    \caption{Spatio-temporal resolution study of $L^2$ error for the $n=5$ perturbation.
    Resolved simulation is $\epsilon_n^{\rm T}$ with $\Delta t = 0.0012$ and $\Delta x = 0.00049$. 
    $R_o = \SI{E2}{\micro\meter}$, $\gamma = \SI{56}{\milli\newton\per\meter}$, $G = \SI{1}{\kilo\pascal}$, $\mu = \SI{10}{\milli\pascal\second}$.
    }
    \label{fig:convergence_study}
\end{figure}

We conduct a resolution study to determine reasonable spatial and temporal discretisation choices.
The initial conditions are for that of a freely oscillating nonspherical bubble $R = R_o$, $\dot{R} = 0$, $T(r,t) = 0$, $\partial T/\partial t = 0$, $\epsilon_5 = 0.1$, $\dot{\epsilon}_5 = 0$.
The parameters are approximately in the middle of the parameter space of simulations in the manuscript: $n = 5$, $R_o = 100 \, \mu$m $\gamma = 56$ mN/m, $G = 1$ kPa, $\mu = 10$ mPa s.
\Cref{fig:convergence_study} shows the $L^2$ norm of the perturbation amplitude difference from the reference resolved simulation versus $\Delta t$ for different $\Delta x$.
Expected second order scaling of the solution is observed.
The error deviates from this scaling as soon as $\Delta x > \Delta t$ (where each individual line plateaus as $\Delta t$ decreases.).
For ultrasound-forced simulations, the end time of each simulation is chosen by $25$ perturbation oscillation cycles or $50$ ultrasound cycles and $\Delta t$ changes between simulations.
However, truncation errors for the simulations still plateau for $\Delta x=0.0078$ and $\Delta t = 0.016$.
Thus, unless stated otherwise, $\Delta x = 0.0078$ and $\Delta t = 0.0118$ ($\SI{E4}{}$ time steps) for the present simulations.

\section{Boundary-layer approximation \label{sec:BL_approx}}

\begin{figure}
    \centering
    \includegraphics[width=1\linewidth]{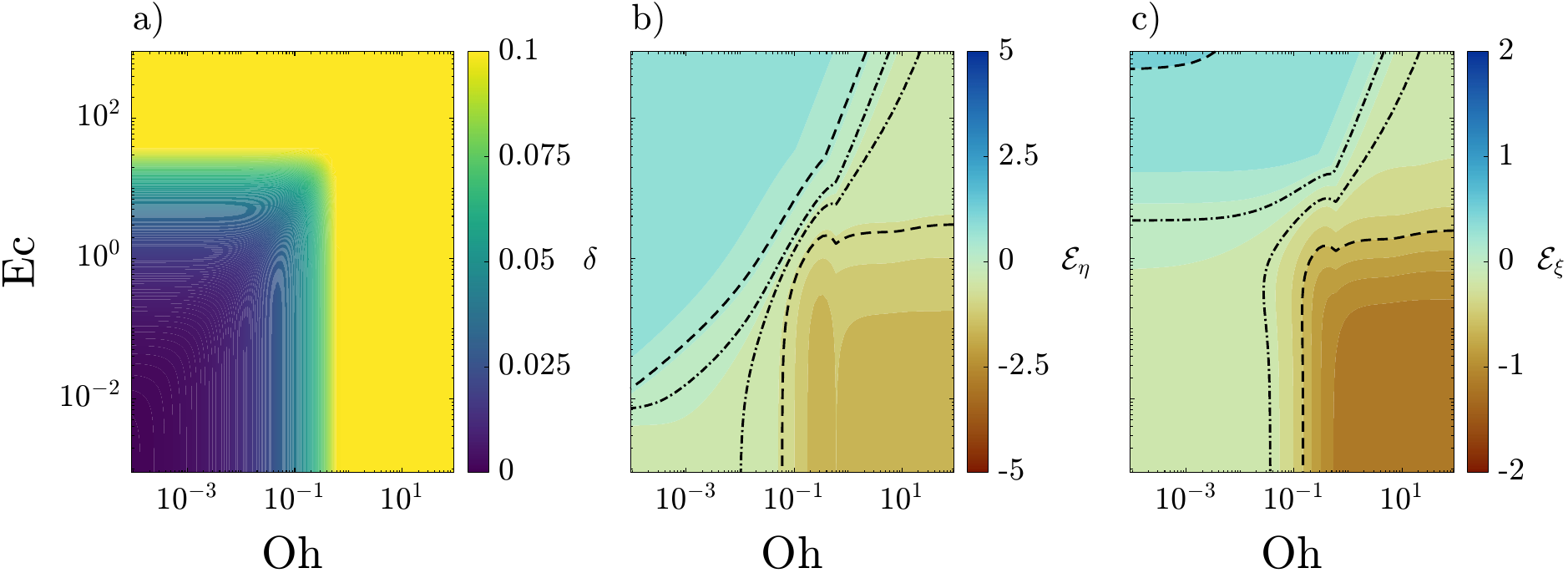}
    \caption{Contours of $\delta$ (a), $\mathcal{E}_\eta$ (b), and $\mathcal{E}_\xi$ (c) for the present boundary layer model.
    Black dashed: $|\mathcal{E}_{(\cdot)}|= 0.5$, dashed-dotted lines: $|\mathcal{E}_{(\cdot)}|= 0.1$.
    Red stars: $\Ec$ and $\Oh$ values of ultrasound forced experiments from \cite{Remillard2026}.
    $n=5$, $R_o = \SI{100}{\micro\meter}$, $\We = 90.5$.}
    \label{fig:potential_comp_VE_BL}
\end{figure}

We compare the boundary layer thickness model from \cref{tab:models} to the full model for the same simulations as in \cref{fig:potential_comp_VE,fig:potential_comp_VE_lamb}.
\Cref{fig:potential_comp_VE_BL} shows the approximation to the boundary layer thickness from \cref{eq:viscoelastic_del_cut}, $\mathcal{E}_{\eta}$, and $\mathcal{E}_{\xi}$ versus $\Oh$ and $\Ec$.
The boundary layer, computed by \cref{eq:viscoelastic_del_cut} is linearly proportional to both $\Ec$ and $\Oh$ and saturates to the geometric cut-off value of $1/(2n)$ for $\Oh > \SI{0.6}{}$ and $\Ec > \SI{40}{}$.
The boundary layer model difference errors moderately match those of \cref{fig:potential_comp_VE_lamb}.

\section{Perturbation amplitude $L^2$ difference\label{sec:L2error}}

\begin{figure}
    \centering
    \includegraphics[width=\linewidth]{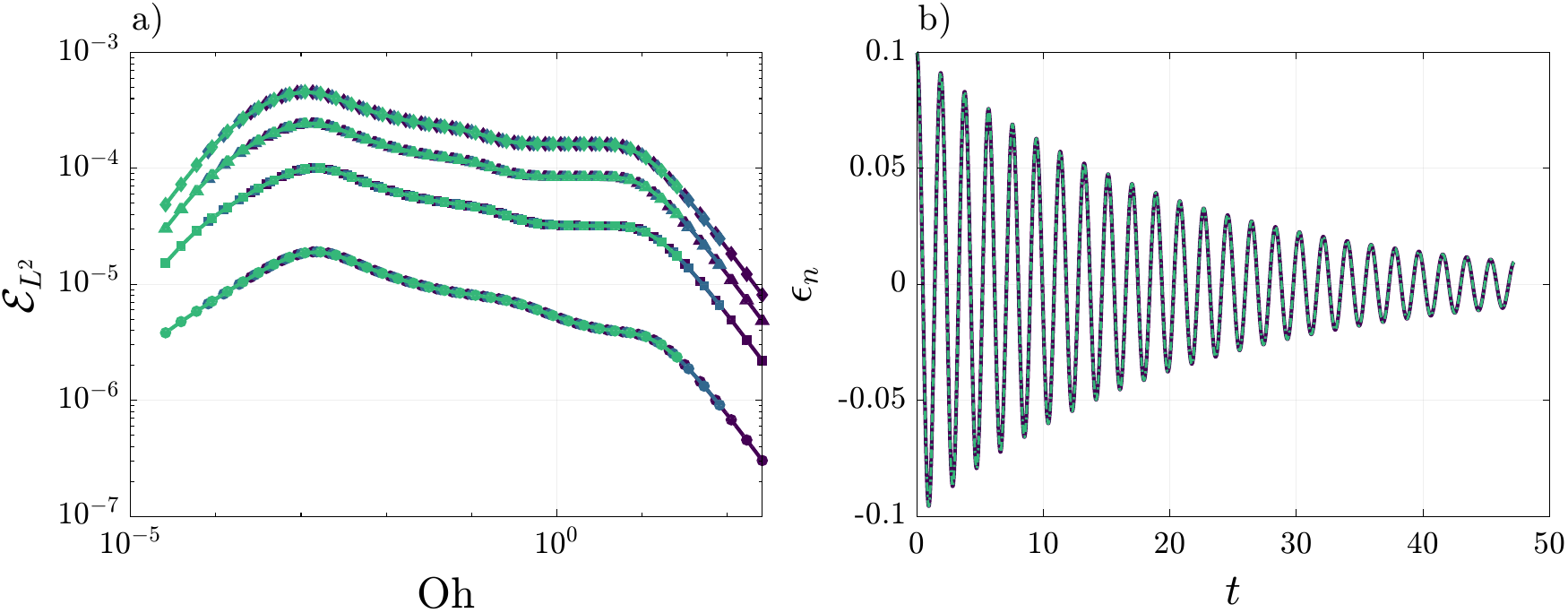}
    \caption{(a) $L^2$ relative difference between the full model and the \cite{Prosperetti1977} model versus $\Oh$. 
    Dark purple $\We = 7.04$, slate blue $\We = 70.4$, olive $\We = 704$. 
    Circles: $n = 2$, squares: $n=5$, triangles: $n = 8$, diamonds: $n = 11$;
    (b) Perturbation with the largest relative difference versus time.
    $\Oh =0.0015$, $\We = 70.4$, $\Rey = 5663$, and $n = 11$.
    Purple solid: present theory, green dashed: \cite{Prosperetti1977}.
    }
    \label{fig:Mod_comp}
\end{figure}

\Cref{fig:Mod_comp} shows the $L^2$ relative difference $\mathcal{E}_{L^2}$ between the full model and the model from \cite{Prosperetti1977} versus $\Oh$.
The relative difference curves are independent of $\We$.
The relative differences are below $\SI{E-3}{}$, when the $L^2$ relative difference is maximised (see frame b).
For small $\Oh$, the models approach the approximately irrotational case.
For large $\Oh$, viscous effects dominate and delay the formation of rotations, and early time $\epsilon_n$ agreement between the models leads to a small $\mathcal{E}_{L^2}$ which does not represent differences in model prediction.

\section{Potential-based model shear strain\label{sec:potential_ss_predict}}
\begin{figure}
    \centering
    \includegraphics[width=\linewidth]{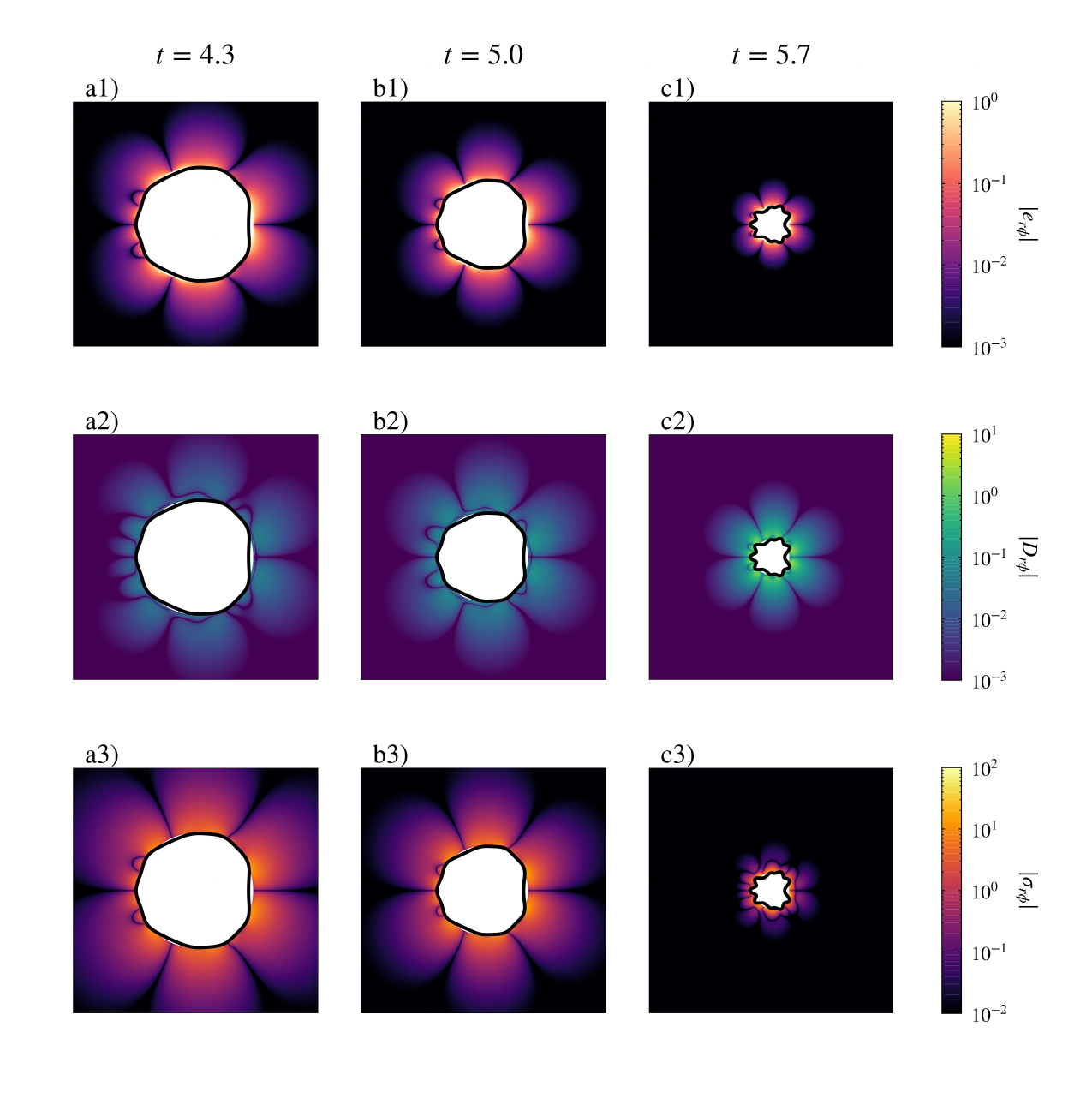}
    \caption{Contours of the Eulerian-Almansi shear strain $e_{r\phi}$ (row 1),  shear strain rate $D_{r\phi}$ (row 2), and shear stress $\sigma_{r\phi}$ (row 3) from the potential-based model from \cite{Remillard2026} with contributions from modes $n=3$, $4$, $6$, $7$, $9$, $10$, and $12$. 
    Material properties \citep{Remillard2026}: $\mu = \SI{0.24}{\pascal \second}$, $G=\SI{2.77}{\kilo\pascal}$, $\gamma=\SI{56}{\milli\newton\per\meter}$, and $\alpha=0$.
    $\lambda_{\rm max} = 5.88$, $R_o=\SI{57.9}{\micro\meter}$, $\Ca = 36.6$, $\We = 52.2$, $\Rey=2.5$.    
    }
    \label{fig:LIC_strain_irr}
\end{figure}

\Cref{fig:LIC_strain_irr} shows contours of the potential-based model shear strain for the LIC initial conditions from \cite{Remillard2026}.
Unlike the full model (see \cref{fig:LIC}), the shear strain is localised to the bubble surface
Additionally, it does not advect radially and exhibits less shear strain and, thus, shear stress.

\bibliographystyle{jfm}
\bibliography{references}

\end{document}


\section{Derivation of $\Phi_n$}
From \citet{Prosperetti1977}, the general solution for the outer fluid:
\begin{equation}
\begin{aligned}
\Phi_n(r,t)
  &=
    \Bigg[
      \left(
        \kappa_n(t)
        + \frac{n+1}{2n+1}
          \int_R^{r} s^{-n} T_n(s,t)\,ds
      \right) r^{n}
      \\
      &+
      \left(
        \frac{n}{n+1} R^{2n+1}\kappa_n(t)
        + \frac{n}{2n+1}
          \int_R^{r} s^{n+1} T_n(s,t)\,ds
      \right) r^{-(n+1)}
    \Bigg] .
    \end{aligned}
\label{eq:prosperetti-18}
\end{equation}

\begin{equation}
\kappa_n(t)
  = -\,\frac{n+1}{2n+1}
    \int_{R}^{\infty} s^{-n} T_n(s,t)\,ds .
\label{eq:prosperetti-19b}
\end{equation}
The solution simplifies to 
\begin{equation}
    \Phi_n(r,t)= \frac{n+1}{2n+1}\mathcal{K}_n(r,t)r^{n}+\left(\frac{n}{n+1} R^{2n+1}\kappa_n(t)+ \frac{n}{2n+1}\mathcal{J}_n(r,t)\right) r^{-(n+1)},
\end{equation}
where
\begin{subequations}
    \begin{equation}
         \mathcal{J}_n(r,t) = \int_R^r s^{n+1} T_n(s,t) d s, 
    \end{equation}
    \begin{equation}
        \frac{\partial \mathcal{J}_n}{\partial r} = r^{n+1}T_n(r,t),
    \end{equation}
    \begin{equation}
        \int_R^r  s^{-n} T_n(s,t) d s = -\frac{2n+1}{n+1}\kappa_n(t) +\mathcal{K}_n(r,t),
    \end{equation}
    \begin{equation}
        \mathcal{K}_n(r,t)=-\int_r^\infty s^{-n}T_n(s,t)ds,
    \end{equation}
    \begin{equation}
        \frac{\partial \mathcal{K}_n}{\partial r} = r^{-n}T_n(r,t),
    \end{equation}
    \begin{equation}
        \frac{\partial \mathcal{J}_n}{\partial t} = \int_R^rs^{n+1}\frac{\partial T_n}{\partial t}ds-\dot{R}R^{n+1}T_n(R,t),
    \end{equation}
    \begin{equation}
         \frac{\partial \mathcal{K}_n}{\partial t} = -\int_r^\infty s^{-n}\frac{\partial T_n}{\partial t}ds,
    \end{equation}
\end{subequations}
Evaluating $\Phi_n$ and its derivatives at the bubble wall: 
\begin{subequations}
    \begin{equation}
        \Phi_n(R,t)=  \frac{2n+1}{n+1}R^n\kappa_n(t),
    \end{equation}
    \begin{equation}
        \left.\frac{\partial\Phi_n}{\partial r}\right|_{R} =  T_n(R,t),
    \end{equation}
    \begin{equation}
    \begin{aligned}
        \left.\frac{\partial\Phi_n}{\partial t}\right|_{R} &= - \dot{R}T_n(r,t)+ \frac{2n+1}{n+1}R^{n-1} \left(n \kappa_n (t) \dot{R}+R\dot{\kappa}_n (t)\right),
    \end{aligned}
    \end{equation}
    \begin{equation}
        \begin{aligned}
         \left.\frac{\partial^2\Phi_n}{\partial t\partial r}\right|_{R} &=  \left.\frac{\partial T_n}{\partial t}\right|_{R}-n (2 n+1) \kappa_n (t) R^{n-2} \dot{R}.
    \end{aligned}
    \end{equation}
\end{subequations}

\section{Dispersion relation}
We use the normal wave ansatz for the the toroidal field $T_n(r,t)= \hat{T_n}e^{i(kr-\Omega t)}$ to find the dispersion relation. 
To get an analytical solution for $\Omega = \Omega(k)$, we neglect the $\Phi_n$ and $\epsilon_n$ equation contributions.
The dispersion relationship is then,
\begin{equation}
    \begin{aligned}
        \Omega &= \frac{(k r+3 i) R^2 \dot{R}}{r^3}-\frac{i \mu  \left(k^2 r^2+n^2+n\right)}{2 \rho r^2} \pm \frac{1}{2 \varrho ^2 \rho r^5}\Bigg[ 4 G \rho \Big((1-3 \alpha ) k^2 \varrho ^8 r^6\\
        &+\alpha  k^2 \varrho ^6 r^2 \left(\varsigma ^2+3 r^6-2 \varsigma  r^3\right)+8 i (3 \alpha -1) \varsigma  k \varrho ^5 r^5-8 i \alpha  \varsigma  k \varrho ^3 r \left(2 \varsigma ^2+3 r^6-4 \varsigma  r^3\right)\\
        &+\alpha  \Big(2 \varsigma ^4 \left(n^2+n-43\right)-8 \varsigma ^3 \left(n^2+n-27\right) r^3+3 n (n+1) r^{12}-2 \varsigma  (n-5) (n+6) r^9\\
        &+\varsigma ^2 (9 n (n+1)-190) r^6\Big)-(3 \alpha -1) \varrho ^2 r^4 \left(-26 \varsigma ^2+n (n+1) r^6+20 \varsigma  r^3\right)\Big)\\
        &-\mu ^2 \varrho ^4 r^6 \left(k^2 r^2+n^2+n\right)^2+4 \varrho ^4 \rho r^4 R \Big(\dot{R} \Big(\mu  r R (-k r (5 k r+36 i)+5 n (n+1)+96)\\
        &+\rho \dot{R} \left(2 i r^3 (k r+4 i)+(17-2 i k r) R^3\right)\Big)+\rho r^3 (-4+i k r) R \ddot{R}\Big)\Bigg]^{1/2}.
    \end{aligned}
\end{equation}
For small radial oscillations, it simplifies to 
\begin{equation}
\begin{aligned}
    \left.\Omega\right|_{R = R_o, \dot{R}, \,  \ddot{R} = 0} &= \frac{1}{2 r^2}\Bigg[-i\left(k^2r^2 + n^2+n \right)\frac{\mu}{\rho} \\
    &\pm\sqrt{\left( k^2r^2+n^2+n\right)\left( 4r^2\frac{G}{\rho}-\frac{\mu^2}{\rho^2}\left( k^2r^2+n^2+n\right)\right)}\Bigg].
\end{aligned}
\end{equation}

\section{Approximately irrotational model for infinitesimally small radial oscillations}

We derive the infinitesimally small radial oscillation reduction of the rotationally-corrected perturbation model and recover Lamb's damping coefficient in the Newtonian limit. 
A thin-boundary-layer closure is applied to the remaining history integral.

\subsection{Governing equations for a static mean radius}
In the limit that $R(t)\approx R_o$, $\dot R \approx 0$, and $\ddot R\approx0$, the bulk equation for the toroidal field becomes
\begin{equation}
\frac{\partial^2 T_n}{\partial t^2}-\frac{\partial^2}{\partial r^2}\left(\frac{\mu}{\rho}\frac{\partial T_n}{\partial t}+\frac{G}{\rho}T_n\right)
+\frac{n(n+1)}{r^2}\left(\frac{\mu}{\rho}\frac{\partial T_n}{\partial t}+\frac{G}{\rho}T_n\right)=0.
\label{eq:bulkT}
\end{equation}
The equation can be written as
\begin{equation}
    \frac{\partial^2 T_n}{\partial t^2}=\frac{1}{\rho}\left(Q_{rr}-\frac{n(n+1)}{r^2}Q\right),
\label{eq:bulkQ}
\end{equation}
where
\begin{subequations}
    \begin{equation}
        Q(r,t)\equiv G T_n(r,t)+\mu\,\frac{\partial T_n}{\partial t}(r,t),
        \label{eq:fancyQ}
    \end{equation}
    \begin{equation}
        \mathcal{X}(t)\equiv G\kappa_n(t)+\mu\dot\kappa_n(t),
        \label{eq:fancyX}
    \end{equation}
    \begin{equation}
        \kappa_n(t)= -\frac{n+1}{2n+1}\int_{R_o}^{\infty} s^{-n}T_n(s,t)\,ds,
    \end{equation}%
    \label{eq:defQchi}%
\end{subequations}%
We can then recast~\cref{eq:fancyX} using~\cref{eq:fancyQ}:
\begin{equation}
    \mathcal{X}(t)= -\frac{n+1}{2n+1}\int_{R_o}^{\infty} s^{-n}Q(s,t)\,ds.
    \label{eq:chiint}
\end{equation}
Plugging \cref{eq:chiint,eq:defQchi} into tangential stress continuity for linearised radial oscillations requires that
\begin{equation}
    \frac{Q(R_o,t)}{R_o} = \frac{2}{n+1}\Big((2n+1)R_o^{n-2}\mathcal{X}(t)+(n+2)(G\epsilon_n+\mu\dot\epsilon_n)\Big).
\label{eq:bcQ}
\end{equation}
The governing equation for the surface perturbation amplitude evolution then simplifies to:
\begin{equation}
\begin{aligned}
\ddot\epsilon_n+\eta\dot\epsilon_n+\xi\epsilon_n
&=
-\frac{2n(n+1)(2n+1)R_o^{n-4}}{\rho}\mathcal{X}
-nR_o^{n-2}\ddot\kappa_n(t) \\
&\quad+
\frac{n(n+1)^2}{R_o^2}
\int_{R_o}^{\infty}
\left[
-\frac{Q}{\rho r^2}
+\frac{1}{2n+1}\left(r^{n-1}\ddot{\mathcal{K}_n}-r^{-(n+2)}\ddot {\mathcal{J}_n}\right)
\right]dr,
\label{eq:epsstart}
\end{aligned}
\end{equation}
with,
\begin{equation}
\mathcal{J}_n(r,t)=\int_{R_o}^{r}s^{n+1}T_n(s,t)\,ds,
\qquad
\mathcal{K}_n(r,t)=-\int_{r}^{\infty}s^{-n}T_n(s,t)\,ds.
\label{eq:JKdefs}
\end{equation}

\subsection{Elimination of nonlocal $J$ and $K$ terms}
Since the integration bounds are fixed,
\begin{subequations}
    \begin{equation}
        \ddot{\mathcal{J}_n}(r,t)=\int_{R_o}^{r}s^{n+1}\frac{\partial^2 T_n}{\partial t^2}(s,t)\,ds,
    \end{equation}
    \begin{equation}
        \ddot{ \mathcal{K}_n}(r,t)=-\int_{r}^{\infty}s^{-n}\frac{\partial^2 T_n}{\partial t^2}(s,t)\,ds.
    \end{equation}
\end{subequations}
Applying Fubini's theorem gives
\begin{align}
\frac{n(n+1)^2}{R_o^2(2n+1)}
\int_{R_o}^{\infty}\left(r^{n-1}\ddot K-r^{-(n+2)}\ddot J\right)dr
&=
-(n+1)R_o^{n-2}\ddot\kappa_n
-\frac{n+1}{R_o^2}\int_{R_o}^{\infty}\frac{\partial^2 T_n}{\partial t^2}\,dr.
\label{eq:JKcollapse}
\end{align}
Substituting \cref{eq:JKcollapse} into \cref{eq:epsstart} yields
\begin{align}
\ddot\epsilon_n+\eta\dot\epsilon_n+\xi\epsilon_n
&=
-\frac{2n(n+1)(2n+1)R_o^{n-4}}{\rho}\mathcal{X}
-(2n+1)R_o^{n-2}\ddot\kappa_n \\
&\quad-
\frac{n(n+1)^2}{\rho R_o^2}\int_{R_o}^{\infty}r^{-2}Q\,dr
-\frac{n+1}{R_o^2}\int_{R_o}^{\infty}\frac{\partial^2 T_n}{\partial t^2}\,dr.
\label{eq:epsmid1}
\end{align}
Then, integrating \cref{eq:bulkQ} from the bubble wall to infinity:
\begin{equation}
\int_{R_o}^{\infty}\frac{\partial^2 T_n}{\partial t^2}\,dr
=
-\frac{1}{\rho}Q_r(R_o,t)-\frac{n(n+1)}{\rho}\int_{R_o}^{\infty}r^{-2}Q\,dr.
\label{eq:intTddot}
\end{equation}
Then, after multiple integration by parts:
\begin{equation}
\begin{aligned}
    \ddot\kappa_n &=-\frac{n+1}{2n+1}\int_{R_o}^{\infty}s^{-n}\frac{\partial^2 T_n}{\partial t^2}(s,t)\,ds, \\
    &=
    \frac{n+1}{(2n+1)\rho}\left(\frac{Q_r(R_o,t)}{R_o^n}+\frac{nQ(R_o,t)}{R_o^{n+1}}\right).
    \label{eq:kappaddotQ}
\end{aligned}
\end{equation}
Substituting \cref{eq:intTddot,eq:kappaddotQ} into \cref{eq:epsmid1}, the $Q_r(R_o,t)$ and $\int r^{-2}Q\,dr$ terms cancel. 
The resulting equation then becomes
\begin{equation}
    \ddot\epsilon_n+\eta\dot\epsilon_n+\xi\epsilon_n = -\frac{2n(n+1)(2n+1)R_o^{n-4}}{\rho}\mathcal{X} -\frac{n(n+1)}{\rho R_o^3}Q(R_o,t).
\label{eq:epsmid2}
\end{equation}
Thus, the full nonlocal structure becomes a pure interfacial rotational correction.

\subsection{First elimination using the interfacial condition}
Rearranging \cref{eq:bcQ} as
\begin{equation}
    (2n+1)R_o^{n-2}\mathcal{X} = \frac{n+1}{2R_o}Q(R_o,t)-(n+2)(G\epsilon_n+\mu\dot\epsilon_n).
\label{eq:chiFromQ}
\end{equation}
Substituting \cref{eq:chiFromQ} into \cref{eq:epsmid2} gives
\begin{equation}
\begin{aligned}
    \ddot\epsilon_n +\left(\eta-\frac{2n(n+1)(n+2)\mu}{\rho R_o^2}\right)\dot\epsilon_n 
    &+\left(\xi-\frac{2n(n+1)(n+2)G}{\rho R_o^2}\right)\epsilon_n\\
    &+\frac{n(n+1)(n+2)}{\rho R_o^3}Q(R_o,t)=0.
\end{aligned}
\label{eq:preelim}
\end{equation}

\subsection{Final simplification using boundary condition for $Q$}
Rewrite \cref{eq:bcQ} as
\begin{equation}
\frac{Q(R_o,t)}{R_o}
=
\frac{2(n+2)}{n+1}(G\epsilon_n+\mu\dot\epsilon_n)
-2R_o^{n-2}\int_{R_o}^{\infty}s^{-n}Q(s,t)\,ds.
\label{eq:QFromchi}
\end{equation}
Substituting \cref{eq:QFromchi} into \cref{eq:preelim} yields
\begin{equation}
\begin{aligned}
\ddot\epsilon_n
+\left(\eta+\frac{2n(n+2)\mu}{\rho R_o^2}\right)\dot\epsilon_n
&+\left(\xi+\frac{2n(n+2)G}{\rho R_o^2}\right)\epsilon_n
\\
&-\frac{2n(n+1)(n+2)R_o^{n-4}}{\rho}\int_{R_o}^{\infty}s^{-n}Q(s,t)\,ds=0.
\end{aligned}
\label{eq:generalPostElim}
\end{equation}
Using the potential-based model coefficients:
\begin{equation}
\eta=\frac{2(n+1)(n+2)\mu}{\rho R_o^2},
\qquad
\xi=\frac{2(n+1)(n+2)}{\rho R_o^2}
\left(G+\frac{(n-1)\gamma}{2R_o}\right),
\label{eq:etaxiPotential}
\end{equation}
\cref{eq:generalPostElim} becomes
\begin{equation}
\begin{aligned}
\ddot\epsilon_n
+\frac{2(n+2)(2n+1)\mu}{\rho R_o^2}\dot\epsilon_n
&+\left[
\frac{2(n+2)(2n+1)G}{\rho R_o^2}
+\frac{(n-1)(n+1)(n+2)\gamma}{\rho R_o^3}
\right]\epsilon_n
\\
&-\frac{2n(n+1)(n+2)R_o^{n-4}}{\rho}\int_{R_o}^{\infty}s^{-n}Q(s,t)\,ds=0.
\end{aligned}
\label{eq:generalizedLambWithHistory}
\end{equation}

If the rotational field is confined to a thin layer near the interface so that the final history
integral is negligible, then \cref{eq:generalizedLambWithHistory} gives the closed ODE
\begin{equation}
\ddot\epsilon_n+\eta_L\dot\epsilon_n+\xi_L\epsilon_n=0,
\end{equation}
with
\begin{equation}
\eta_L=\frac{2(n+2)(2n+1)\mu}{\rho R_o^2},
\qquad
\xi_L=\frac{2(n+2)(2n+1)G}{\rho R_o^2}
+\frac{(n-1)(n+1)(n+2)\gamma}{\rho R_o^3}.
\end{equation}
Thus, the corrected constant-radius reduction recovers Lamb's damping coefficient exactly.

\subsection{Thin-boundary-layer approximation for the final history integral}
We now apply a boundary layer approximation to the last integral in
\cref{eq:generalizedLambWithHistory}. 
Let the rotational traction variable be concentrated in a layer of thickness, i.e.,
\begin{equation}
    \ell=\delta R_o,
\end{equation}
where $0<\delta\ll 1$ and $Q(r,t)$ approximately equal to its interfacial value across the layer. 
Then the integral on the right hand side of \cref{eq:generalizedLambWithHistory} becomes,
\begin{equation}
    \int_{R_o}^{\infty}s^{-n}Q(s,t)\,ds \approx
    Q(R_o,t)\int_{R_o}^{R_o+\ell}s^{-n}\,ds \approx \delta R_o^{1-n}Q(R_o,t).
    \label{eq:BLint1}
\end{equation}
Substituting \cref{eq:BLint1} into the boundary condition \cref{eq:QFromchi} gives
\begin{equation}
    \frac{Q(R_o,t)}{R_o} \approx \frac{2(n+2)}{n+1}(G\epsilon_n+\mu\dot\epsilon_n)-2\delta\frac{Q(R_o,t)}{R_o},
\end{equation}
and therefore
\begin{equation}
    Q(R_o,t) \approx \frac{2(n+2)R_o}{(n+1)(1+2\delta)}(G\epsilon_n+\mu\dot\epsilon_n).
\label{eq:QwallBL}
\end{equation}
Using \cref{eq:BLint1} and \cref{eq:QwallBL}, the history integral becomes
\begin{equation}
    \int_{R_o}^{\infty}s^{-n}Q(s,t)\,ds \approx \frac{2(n+2)\delta}{(n+1)(1+2\delta)}R_o^{2-n}(G\epsilon_n+\mu\dot\epsilon_n).
\label{eq:BLint2}
\end{equation}
Substituting \cref{eq:BLint2} into \cref{eq:generalizedLambWithHistory} yields the closed
boundary-layer approximation
\begin{equation}
    \ddot\epsilon_n+\eta_{\mathrm{BL}}\dot\epsilon_n+\xi_{\mathrm{BL}}\epsilon_n=0,
\label{eq:BLfinalODE}
\end{equation}
with effective coefficients

\begin{align}
\eta_{\mathrm{BL}}
&=
\frac{2(n+2)(2n+1)\mu}{\rho R_o^2}
-\frac{4n(n+2)^2\delta}{(1+2\delta)}\frac{\mu}{\rho R_o^2},
\label{eq:etaBL}\\
\xi_{\mathrm{BL}}
&=
\frac{2(n+2)(2n+1)G}{\rho R_o^2}
+\frac{(n-1)(n+1)(n+2)\gamma}{\rho R_o^3}
-\frac{4n(n+2)^2\delta}{(1+2\delta)}\frac{G}{\rho R_o^2}.
\label{eq:xiBL}
\end{align}
Equivalently,
\begin{align}
\eta_{\mathrm{BL}}
&=
\frac{2(n+2)\mu}{\rho R_o^2}
\left[(2n+1)-\frac{2n(n+2)\delta}{1+2\delta}\right],\\
\xi_{\mathrm{BL}}
&=
\frac{(n-1)(n+1)(n+2)\gamma}{\rho R_o^3}
+\frac{2(n+2)G}{\rho R_o^2}
\left[(2n+1)-\frac{2n(n+2)\delta}{1+2\delta}\right].
\end{align}
The modified damping coefficient matches that of \cite{Hilgenfeldt1996}.

\bibliographystyle{jfm}
\bibliography{references}